\documentclass[dvips]{article}
\usepackage{graphicx}
\usepackage{amssymb}
\usepackage{amsfonts}
\usepackage[bookmarks=false]{hyperref}

\evensidemargin=\oddsidemargin
\newdimen\dummy
\dummy=\oddsidemargin
\newtheorem{theorem}{Theorem}

\newtheorem{lemma}[theorem]{Lemma}

\newenvironment{proof}[1][Proof]{\noindent\textbf{#1.} }{\ \rule{0.5em}{0.5em}}

\begin{document}

\title{Single Photon Ignition of Two-photon Super-fluorescence through the
Vacuum of Electromagnetic Field }
\author{Nicolae A. Enaki \\
Academiei str.5, Institute of Applied Physics of Academy of sciences of
Moldova,\\
Chisinau, MD 2028, Republic of Moldova}
\date{December 8, 2010 }
\maketitle

\begin{abstract}
The ignition of two-quantum collective emission of inverted sub-ensemble of
radiators due to mutual interaction of this sub-ensemble with other two
dipole active atomic subsystems in process of two-photon exchanges between
the atoms through the vacuum field is proposed. The three particle
resonances between two-photon and single quantum transitions of inverted
radiators from the ensemble are proposed for acceleration of collective
decay rate of bi-photons, obtained relatively dipole-forbidden transitions
of excited atomic sub-ensemble. This mutual interaction between three
super-fluorescent processes in subatomic ensembles take place relatively
dipole-forbidden transitions in one of radiator subsystem. The collective
resonance emission and absorption of two-quanta have nontraditional
behavior, accompanied with acceleration and inhibition of collective
emission processes of photons.
\end{abstract}

\section{Introduction}

A great deal of attention is currently devoted to the problem of coherence
which appears not only between the quanta but between groups of quanta too.
The generation of non-classical coherent electromagnetic field in
multi-photon emission and the interaction of coherent radiation with matter
(nuclei, atoms and solids) have been subjects of a number of theoretical and
experimental studies in recent years \cite{NZT_1981}-\cite{BRH_1987}.
Examples include the higher-order coherence in multi-photon generation of
light the two-photon micro-maser emission \cite{BRH_1987}, two-photon lasers
the parametric down conversion, four-wave mixing and other effects in
optical diapason \cite{GWMM_1992}, and the possibility of coherent
generation of photons in $x$\textbf{-}ray and\textbf{\ }$gamma$\textbf{-\ }%
ray spectral regions.

In this article it is proposed to investigate the cooperative two-photon
emission from inverted system of radiators stimulated by single photon
super-fluorescent pulses in two-quantum resonance with dipole forbidden
atomic transition. Since the two-photon cooperative phenomenon has the small
two-photon cooperative emission time \cite{E_1988}, we propose to extend our
attention to the new tape of cooperative resonance interaction between three
radiators in which single photon transitions of two radiators which enter in
two-photon resonance with dipole forbidden transition of third atom. This
cooperative three particle interaction take place through the vacuum
fluctuations of electromagnetic field and can amplify or diminish the
spontaneous emission rates of the atoms. In order to obtain more powerful
pulses of entangled photons it is proposed the cooperative interaction
between three atomic subsystems in which one of them are inverted relatively
dipole forbidden transition {$\vert2S>$ $-$ $\vert S>$ of Hydrogen like or
Helium Like atoms \cite{BT_1940}-\cite{AE_1991}. Taking in to account the
elementary acts of two photon interaction between radiators we archived the
improvement of two-photon emission rate of the system of radiators in
comparison with two-photon super-fluorescence \cite{E_1988}. In this article
it is examined the mutual influence of two single-photon super-fluorescence
processes and two-photon cooperative emission of the atomic system
relatively dipole forbidden transition. The phenomenon of new cooperative
emission takes in to account the three particle mutual interaction with
vacuum of electromagnetic field in which the product of vacuum polarization
of two atoms enter in to resonance with two-photon polarization of dipole
forbidden transition of Hydrogen-like or Helium-like radiator. It has been
shown that in the process of spontaneous radiation, the radiators (nuclei,
atoms) enter a regime of single and two-photon super-radiance and the rate
of photon pair (bi-photon) emission increases (or decreases) due to new
three particle cooperative phenomenon, which appear between single and
two-photon spontaneous emission subgroups of radiators. It has been
demonstrated, that for hydrogen-like and helium-like atoms \cite{MS_1972}
the dipole-forbidden transitions can generate more powerful pulse of
entangled photon pairs (bi-photons) under the influence of single photon
super-radiance. }

It is important to note that for coherent radiation of such system, was
studied for the dimension of a radiating system smaller than the radiation
wavelength. It is, however, interesting to study this type of cooperative
emission between three radiator subsystems in extended system of radiators.
The possibilities of two-photon cooperative resonances between three
radiators replaced at distance larger than emission wavelength are studied
too. I emphasize here, that the problem of cooperation between two single
photon cooperative emission subsystems and one two-photon cooperative
emission subsystem is more complicated than the similar problem of single
\cite{D_1954} or two-photon \cite{E_1988} super-radiances in extended
system. In Dicke's super-radiance, the exchange integral between $j$-th and $%
l$-th atoms is described by more simple exchange integral proportional to $%
\sin [k_{0}r_{jl}]/(k_{0}r_{jl})$ while in two-photon super-radiance by more
complicated function $\sin [(2k_{0}-k)r_{jl}]/[(2k_{0}-k)r_{jl}]\times \sin
[kr_{jl}]/(kr_{jl})$, where $r_{jl}$ and $k_{i}\ $are the distance between
the radiators and wave vector of emitted photons respectively. The three
particle exchange integral between $j$-th, $m$-th and $l$-th atoms was
obtained in this paper taking in to account the two-quantum exchanges
between two radiators proposed in papers \cite{E_1988}, \cite{EM_1997}. The
more complicated exchange integrals between two radiators with dipole active
transition and one radiator with dipole forbidden transition is given in
Appendix of this paper.

\section{Interaction Hamiltonian and Master Equation}

Let us consider the interaction of three subsystems of radiators $R$, $S$,
and $D$ thorough vacuum of electromagnetic field. The first two groups, $R$
and $S$, are prepared in excited state $|e_{r}\rangle \otimes |e_{s}\rangle $
and can pass in to Decke super-radiance regime \cite{D_1954} relatively the
dipole active transitions $e_{r}\rightarrow g_{r}$ and $e_{s}\rightarrow
g_{s}$ at frequencies $\omega _{r}$ and $\omega _{s}$ (see figure \ref%
{figure_1}). The $D$ atomic subsystem is prepared in excited state $%
|e_{d}\rangle $ and relatively dipole forbidden transition $e_{d}\rightarrow
g_{d}$ and can pass in the ground state $|g_{d}\rangle $ \ simultaneously
generation two quanta \cite{E_1988}. Let us consider the simple cooperative
stimulation of two-photon emission of $D$ system stimulated by $R$ and $S$
radiator subsystems.

\begin{figure}[t]
\begin{center}
\includegraphics[width=10cm,height=6cm]{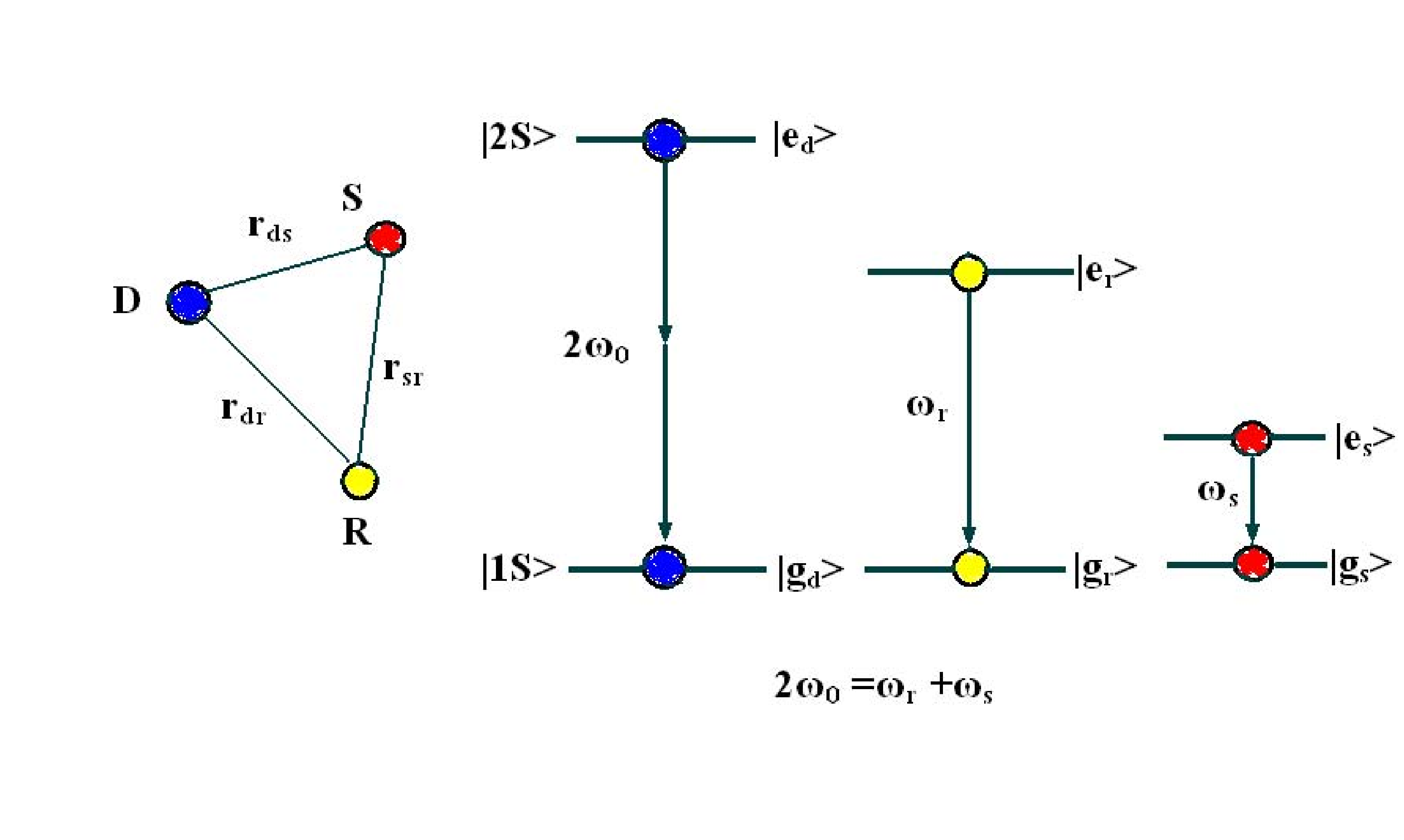}
\end{center}
\caption{The resonance between two-photon transitions of D atomic subsystem
and two dipole active atomic subsystems$R$ and $S$ . As an example is
represented three atoms $D$, $R$ and $S$ situated at relatively distances $%
r_{ds}$, $r_{dr}$ and $r_{rs}$. One of condition of exchange energies
between the subsystems is the e resonance between two-photon and single
photon transitions $2\protect\omega_0 =\protect\omega_r+\protect\omega_s$ .}
\label{figure_1}
\end{figure}

In this case it is established the resonance between the dipole active
atomic subgroups $R$, $S$ and dipole forbidden radiators of $D$ ensemble and
the cooperative stimulation of two-quantum collective transition is
possible. Indeed, considering that the conservation energy law is
established between these groups, $\hbar (\omega _{r}+\omega _{s})=2\hbar
\omega _{0}$, one can proposed the following Hamiltonian of interaction of
radiators with electromagnetic field
\begin{eqnarray}
H &=&H_{0}+\lambda H_{I};  \nonumber \\
H_{0} &=&\sum\limits_{k}\hbar \omega _{k}a_{k}^{\dagger
}a_{k}+\sum\limits_{j=1}^{N_{r}}\hbar \omega
_{r}R_{zj}+\sum\limits_{l=1}^{N_{s}}\hbar \omega
_{s}S_{zl}+2\sum\limits_{m=1}^{N}\hbar \omega _{0}D_{zm};  \nonumber \\
\lambda H_{I} &=&-\sum\limits_{k}\sum\limits_{j=1}^{N_{a}}(\mathbf{d}_{r},%
\mathbf{g}_{k})\{R_{j}^{+}a_{k}\exp [i(\mathbf{k},\mathbf{r}%
_{j})]+R_{j}^{-}a_{k}^{\dagger }\exp [-i(\mathbf{k},\mathbf{r}_{j})]\}
\nonumber \\
&-&\sum\limits_{k}\sum\limits_{l=1}^{N_{b}}(\mathbf{d}_{s},\mathbf{g}%
_{k})\{S_{l}^{+}a_{k}\exp [i(\mathbf{k},\mathbf{r}_{l})]+S_{l}^{-}a_{k}^{%
\dagger }\exp [-i(\mathbf{k},\mathbf{r}_{l})]\}  \nonumber \\
&-&\sum\limits_{k_{1},k_{2}}\sum\limits_{m=1}^{N_{b}}(\mathbf{n}_{eg},%
\mathbf{e}_{\lambda _{1}})(\mathbf{n}_{ei},\mathbf{e}_{\lambda
_{2}})q(\omega _{1},\omega _{2})  \nonumber \\
&\times &\{D_{m}^{+}a_{k_{2}}a_{k_{1}}\exp [i(\mathbf{k}_{1}+\mathbf{k}_{2},%
\mathbf{r}_{m})]  \nonumber \\
&+&D_{m}^{-}a_{k_{1}}^{\dagger }a_{k_{2}}^{\dagger }\exp [-i(\mathbf{k}_{1}+%
\mathbf{k}_{2},\mathbf{r}_{l})]\}.  \label{eq:H}
\end{eqnarray}%
Here

\[
q(\omega _{1},\omega _{2})=\frac{d_{23}d_{31}g_{k_{2}}g_{k_{1}}}{2\hbar }%
\left\{ \frac{1}{\omega _{32}+\omega _{k_{1}}}+\frac{1}{\omega _{31}-\omega
_{k_{2}}}\right\} ,\ \ \ \mathbf{g}_{k}=\sqrt{\frac{2\pi \hbar \omega _{k}}{V%
}}\mathbf{\epsilon }_{_{\lambda }},
\]%
$a_{k}$ and $a_{k}^{\dagger }$ are annihilation and creation operators of \
EMF photons with wave vector $\mathbf{k,}$ polarization \textbf{\ }$\mathbf{%
\ \epsilon }_{_{\lambda }}$\textbf{\ }and frequency $\omega _{k}$\textbf{;} $%
\mathbf{d}_{r}$ and $\mathbf{d}_{s}$ are dipole momentum transition between
the ground and excited states for $R$ and $S$ atomic subsystems; $d_{ei}$
and $d_{eg}$ are dipole momentum transitions in the three level system of
atomic group $D$. The operators of $R$, $S$, and $D$ atomic subsystems
satisfy the commutation relations for $SU(2)$ algebra $[J^{+},J^{-}]=2J_{z}$%
;\ $[J_{z}\ ,J^{\pm }]=\pm J^{\pm }$,\ where $J^{\pm }$ is equivalent with $%
R^{\pm }$, $S^{\pm }$ and $D^{\pm }$. Invertin operator $J_{z}$ is consider
similar to $R_{z},$ $S_{z}$ and $D_{z}$ respectively. The operators of
electromagnetic field satisfy the commutation relation $[a_{k},a_{k^{\prime
}}^{\dagger }]=\delta _{k,k^{\prime }}\ ;[a_{k}^{\dagger },a_{k^{\prime
}}^{\dagger }]=0$, where $k=(\mathbf{k,}\lambda )$ is the wave vector and
polarization of the photon. Taking in to account the Hamiltonian \ref{eq:H},
let us represent the solution of Haisenberg equation through the sources and
free part operators
\begin{equation}
a_{k}(t)=a_{k}(0)\exp [-i\omega _{k}t]+a_{ks}(t),  \label{eq:as}
\end{equation}%
where the source part is
\begin{eqnarray*}
a_{ks}(t) &=&\frac{i(\mathbf{d}_{a},\mathbf{g}_{k})}{\hbar }%
\sum\limits_{l=1}^{N_{a}}\exp [-i(\mathbf{k},\mathbf{r}_{l})\int%
\limits_{0}^{t}d\tau \exp [-i\omega _{k}\tau ]R_{l}^{-}(t-\tau ) \\
&&+i\frac{(\mathbf{d}_{b},\mathbf{g}_{k})}{\hbar }\sum\limits_{j=1}^{N_{b}}%
\exp [-i(\mathbf{k},\mathbf{r}_{j})\int\limits_{0}^{t}d\tau \exp [-i\omega
_{k}\tau ]S_{j}^{-}(t-\tau ) \\
&&+2i\sum\limits_{n=1}^{N_{b}}\sum\limits_{k_{1}}\frac{(\mathbf{n}_{eg},%
\mathbf{e}_{\lambda _{1}})(\mathbf{n}_{ei},\mathbf{e}_{\lambda })q(\omega
_{1},\omega )}{\hbar }\exp [-i(\mathbf{k}_{1}+\mathbf{k},\mathbf{r}_{n}) \\
\times &&\int\limits_{0}^{t}d\tau \exp [-i\omega _{k}\tau ]D_{n}^{-}(t-\tau
)a_{k_{1}}^{\dagger }(t-\tau );\ \ \ [a_{ks}^{\dagger
}(t)=[a_{s}^{{}}(t)]^{+}.
\end{eqnarray*}%
Taking in to account that $a_{k}(0)|0\rangle _{ph}=\langle
0|_{ph}a_{k}^{\dagger }(0)=0$ we can partially eliminate the EMF field
operators from the mean value of Hesenberg equation for arbitrary atomic
operator $O(t)$
\begin{eqnarray}
\frac{d}{dt}\langle O(t)\rangle =-i\sum\limits_{k}\sum\limits_{j=1}^{N_{a}}%
\frac{(\mathbf{d}_{a},\mathbf{g}_{k})}{\hbar }\langle \lbrack
R_{j}^{+}(t),O(t)]a_{ks}(t)\rangle \exp [i(\mathbf{k},\mathbf{r}_{j})] &&
\nonumber \\
-i\sum\limits_{k}\sum\limits_{l=1}^{N_{b}}\frac{(\mathbf{d}_{b},\mathbf{g}%
_{k})}{\hbar }\langle \lbrack S_{l}^{+}(t),O(t)]a_{ks}(t)\rangle \exp [i(%
\mathbf{k},\mathbf{r}_{l})] &&  \nonumber \\
-i\sum\limits_{k_{1},k}\sum\limits_{m=1}^{N_{b}}\frac{(\mathbf{n}_{eg},%
\mathbf{e}_{\lambda _{1}}(k_{1}))(\mathbf{n}_{ei},\mathbf{e}_{\lambda
}(k))q(\omega _{k_{1}},\omega _{k})}{\hbar } &&  \nonumber \\
\times \langle \lbrack D_{m}^{+}(t),O(t)]a_{k_{1}}(t)a_{ks}(t)\rangle \exp
[i(\mathbf{k}_{1}+\mathbf{k},\mathbf{r}_{m})]+H.C.(O^{+}\rightarrow O). &&
\label{eq:O}
\end{eqnarray}%
Here the mean values of Hesenberg operators are considered taking into
account the initial state of the system $|\Psi _{r}(0)\rangle \otimes
|0\rangle _{ph},$. where $|\Psi _{r}(0)\rangle $ is the state of radiator
subsystem, and $|0\rangle _{ph}$ is the vacuum state of EMF. We are
interested in the total elimination of operators of electromagnetic field
from the expression (\ref{eq:O}) For elimination of operators of
electromagnetic field we formulate the lemma

\begin{lemma}
If Bose $a_{k}(t)$ and $a_{k}^{+}(t)$ operators lie between the two
operators of the atomic subsystem $A(t_{1})$ and $B(t_{2})$ \ ($A(t_{1})$ , $%
B(t_{2})$ don't contain the operators $a_{k}$ and $a_{k}$ ) belonging to
other times, the elimination of the free part of these operators yields the
following expression for the correlation:
\begin{eqnarray}
\left\langle A(t_{1})a_{k}(t)B(t_{2})\right\rangle &=&\left\langle
A(t_{1})a_{ks}(t)B(t_{2})\right\rangle  \nonumber \\
- &&e^{-i\omega _{3}(t-t_{2})}\left\langle
A(t_{1})[a_{ks}(t_{2}),B(t_{2})]\right\rangle ,  \nonumber \\
\left\langle A(t_{1})a_{k}^{+}(t)B(t_{2})\right\rangle &=&\left\langle
A(t_{1})a_{ks}^{+}(t)(t)B(t_{2})\right\rangle  \nonumber \\
&-&e^{i\omega _{3}(t-t_{1})}\left\langle
[A(t_{1}),a_{ks}^{+}(t_{1})]B(t_{2})\right\rangle .  \label{L1}
\end{eqnarray}
\end{lemma}

\begin{proof}
The commutations in (\ref{L1}) play the highest role in the two-photon
spontaneous emission and only such commutations bring the main contribution
to the two-photon process. The problem is reduced to the elimination of
vacuum part lies between the operators $A(t_{1})$ and $B(t_{2})$
\begin{equation}
<A(t_{1})a_{k}(t)B(t_{2})>=<A(t_{1})(a_{k}^{v}(t)+a_{ks}(t))B(t_{2})>
\label{PL11}
\end{equation}%
Since \ $a_{k}(t)=a_{k}^{v}(t)+a_{ks}(t),\;$we will represent the vacuum
part $a_{k}^{v}(t)=a_{k}(0)\exp [-i\omega _{k}t]$ through the
vacuum-operator at time $t_{1\;\ }$and taking in to account the identity (%
\ref{eq:as}) we can represent the vacuum part in the following form $%
a_{k}^{v}(t)=a_{k}^{v}(t_{2})e^{-i\omega
_{k}(t-t_{2})}=\{a_{k}(t_{2})-a_{ks}(t_{2})\}e^{-i\omega _{k}(t-t_{2})}.$
After substitution of $\ a_{k}^{v}(t)\;$into$\;$the correlation it is obtain
\begin{eqnarray*}
\langle A(t_{1})a_{k}(t)B(t_{2})\rangle &=&\langle
A(t_{1})a_{ks}(t)B(t_{2})\rangle \\
&+&e^{-i\omega _{k}(t-t_{2})}\langle
A(t_{1})\{a_{k}(t_{2})-a_{ks}(t_{2})\}B(t_{2})\rangle ,
\end{eqnarray*}%
We observe that $a_{k}(t_{2})\;$commutes with the operator $B(t_{2})$.\
Consequently taking into account that

$a_{k}(t_{2})\ B(t_{2})|0>=\ B(t_{2})a_{ks}(t_{2})|0>,\;$it is easily obtain
that

$\left\langle A(t_{1})\{a_{k}(t_{2})-a_{ks}(t_{2})\}\ B(t_{2})\right\rangle
=-\left\langle B(t_{2})[a_{ks}(t_{2}),B(t_{2})]\right\rangle $.

This relation proofs the Lemma.
\end{proof}

This lemma (\ref{L1}) can be used in the last term of generalized equation (%
\ref{eq:O}) for correlation functions $\ \langle \lbrack
D_{j}^{+}(t),O(t)]a_{k}(t)S_{n}^{-}(t-\tau )\rangle $ , $\ \langle \lbrack
D_{j}^{+}(t),O(t)]a_{k}(t)R_{l}^{-}(t-\tau )\rangle \ $\ and $\langle
\lbrack R_{j}^{+}(t),O(t)]D_{n}^{-}(t-\tau )a_{k_{1}}^{\dagger }(t-\tau
)\rangle ,$ $\langle \lbrack S_{l}^{+}(t),O(t)]D_{n}^{-}(t-\tau
)a_{k_{1}}^{\dagger }(t-\tau )\rangle $. Indeed taking in to account the
lemma (\ref{L1}) the above correlation functions can be represented through
atomic operators%
\begin{eqnarray}
\langle \lbrack D_{j}^{+}(t),O(t)]a_{k}(t)S_{n}^{-}(t-\tau )\rangle
&=&\left\langle [D_{j}^{+}(t),O(t)]a_{ks}(t)S_{n}^{-}(t-\tau )\right\rangle
\nonumber \\
- &&e^{-i\omega _{k}\tau }\left\langle [D_{j}^{+}(t),O(t)][a_{ks}(t-\tau
),S_{n}^{-}(t-\tau )]\right\rangle ,  \label{eq:DRS}
\end{eqnarray}%
\begin{eqnarray}
\langle \lbrack R_{j}^{+}(t),O(t)]a_{k}^{\dagger }(t-\tau )D_{n}^{-}(t-\tau
)\rangle &=&\langle \lbrack R_{j}^{+}(t),O(t)]a_{ks}^{\dagger }(t-\tau
)D_{n}^{-}(t-\tau )\rangle  \nonumber \\
- &&e^{-i\omega _{k}\tau }\langle \lbrack \lbrack
R_{j}^{+}(t),O(t)],a_{ks}^{\dagger }(t)]D_{n}^{-}(t-\tau )\rangle .
\label{eq:RSD}
\end{eqnarray}%
The interaction between the atomic subsystems can be found in the third
order of interaction constants with the subsystems $S$, $R$ and $D$
respectively $(\mathbf{d}_{r},\mathbf{g}_{k})(\mathbf{d}_{s},\mathbf{g}%
_{k})q(\omega _{1},\omega _{2})$ \ According with this condition the smooth
correlation functions is obtained only for the following terms of
expressions (\ref{eq:DRS}) and (\ref{eq:RSD})\ : $\left\langle
R_{l}^{-}(t-\tau ^{\prime })[D_{j}^{+}(t),O(t)]S_{n}^{-}(t-\tau
)\right\rangle $ and $\langle R_{l}^{+}(t-\tau ^{\prime
})[R_{j}^{+}(t),O(t)]D_{n}^{-}(t-\tau )\rangle $. The contribution of other
terms of the expressions (\ref{eq:DRS}) and (\ref{eq:RSD}) give the
contribution more hair order on the decomposition on the small parameter $%
\lambda $ of the interaction Hamiltonian (\ref{eq:H}).

The lemma (\ref{L1}) is non-applicable for correlation functions in which it
is meet simultaneously the creation and annihilations Boson operators
belonging to different time intervals: $\langle \lbrack
D_{j}^{+}(t),O(t)]a_{k}(t)D_{n}^{-}(t-\tau )a_{k_{1}}^{\dagger }(t-\tau
)\rangle $ and its hermit conjugate part $\langle a_{k_{1}}(t-\tau
)D_{n}^{+}(t-\tau )a_{k}^{\dagger }(t)[O(t),D_{j}^{-}(t)]\rangle $. In order
to eliminate the vacuum part of operators $a_{k}(t)$ and $a_{k}^{\dagger
}(t^{\prime })$ let us formulate the following rule.

\begin{lemma}
If the operators $A(t_{1})$ contains the creation operators of EMF and $%
B(t_{2})$ contains the annihilation operators of EMF the elimination of \
vacuum part of annihilation $a_{k}(t)$ or creation $a_{k}^{\dag }(t)$
operators situated between these operators $A(t_{1})$ and $B(t_{2})$ takes
place according with Lemma 1. In opposite case, when operator $A(t_{1})$ can
be represented through the product of atomic operator $\mathcal{A}(t_{1})$
and \ annihilation field operators $A(t_{1})$=$\mathcal{A}%
(t_{1})a_{k_{1}}(t_{1})a_{k_{2}}(t_{1})...a_{k_{n}}(t_{1})$ .\ \ the
operator $B(t_{2})$ is represented through the product of creation field
operators and atomic operator $\mathcal{B}(t_{2})$ so that B(t$_{2}$)=$%
\mathcal{B}(t_{2})a_{k_{1}}^{\dag }(t_{2})a_{k_{2}}^{\dag
}(t_{2})...a_{k_{m}}^{\dag }(t_{2})$ the elimination of vacuum part of the
operators $a_{k}(t)$ and $a_{k}^{\dag }(t)$ can be represented in the
following form%
\begin{eqnarray}
&&\left\langle A(t_{1})a_{k}(t)B(t_{2})\right\rangle =\left\langle
A(t_{1})a_{ks}(t)B(t_{2})\right\rangle  \nonumber \\
&&-\exp [-i\omega _{k}(t-t_{2})]\{\left\langle
A(t_{1})[a_{ks}(t_{2}),B(t_{2})]\right\rangle  \nonumber \\
&&-\delta _{k,k_{1}}\left\langle A(t_{1})\mathcal{B}(t_{2})a_{k_{2}}^{\dag
}(t_{2})...a_{k_{m}}^{\dag }(t_{2})\right\rangle  \nonumber \\
&&-...-\delta _{k,k_{m}}\left\langle A(t_{1})\mathcal{B}(t_{2})a_{k_{1}}^{%
\dag }(t_{2})a_{k_{2}}^{\dag }(t_{2})...a_{k_{m-1}}^{\dag
}(t_{2})\right\rangle \},  \label{L21}
\end{eqnarray}%
\begin{eqnarray}
&&\left\langle A(t_{1})a_{k}^{\dag }(t)B(t_{2})\right\rangle =\left\langle
A(t_{1})a_{ks}^{\dag }(t)B(t_{2})\right\rangle  \nonumber \\
&&-\exp [i\omega _{k}(t-t_{2})]\{\left\langle [A(t_{1})a_{ks}^{\dag
}(t_{1})]B(t_{2})\right\rangle  \nonumber \\
&&-\delta _{k,k_{1}}\left\langle \mathcal{A}%
(t_{1})a_{k_{2}}(t_{1})...a_{k_{n}}(t_{1})B(t_{2})\right\rangle  \nonumber \\
&&-...-\delta _{k,k_{n}}\left\langle \mathcal{A}%
(t_{1})a_{k_{1}}(t_{1})a_{k_{2}}(t_{1})...a_{k_{n-1}}(t_{1})B(t_{2})\right%
\rangle \}.  \label{L22}
\end{eqnarray}
\end{lemma}

\begin{proof}
Taking in to account the lemma (\ref{L1}), we can represent the third
correlation $\left\langle A(t_{1})a_{k}(t)B(t_{2})\right\rangle $ of
expression (\ref{L21}) in the following form

\begin{eqnarray}
\left\langle A(t_{1})a_{k}(t)B(t_{2})\right\rangle &=&\left\langle
A(t_{1})a_{ks}(t)B(t_{2})\right\rangle  \nonumber \\
&+&\exp [-i\omega _{k}(t-t_{2})]\left\langle
A(t_{1})(a_{k}(t_{2})-a_{ks}(t_{2}))B(t_{2})]\right\rangle  \nonumber \\
&-&\exp [-i\omega _{k}(t-t_{2})]\{\left\langle
A(t_{1})[a_{ks}(t_{2}),B(t_{2})]\right\rangle .  \label{PL21}
\end{eqnarray}%
According with explicit expression of operator, $B(t_{2})=\mathcal{B}%
(t_{2})a_{k_{1}}^{\dag }(t_{2})a_{k_{2}}^{\dag }(t_{2})...a_{k_{m}}^{\dag
}(t_{2})$, let us introduced it in the third term of right hand site of
expression (\ref{PL21}). Following the commutation roles of boson operators
of electromagnetic field, the operator $a_{k}(t_{2})$ can be permuted in the
right hand site of the correlation. Taking in to consideration that $%
(a_{ks}(t_{2})+a_{kv}(t_{2}))\left\vert 0\right\rangle
=a_{ks}(t_{2})\left\vert 0\right\rangle ,$ \ this term becomes
\begin{eqnarray}
&&\left\langle A(t_{1})a_{k}(t_{2})\mathcal{B}(t_{2})a_{k_{1}}^{\dag
}(t_{2})a_{k_{2}}^{\dag }(t_{2})...a_{k_{m}}^{\dag }(t_{2})\right\rangle =
\nonumber \\
&&\delta _{k,k_{1}}\left\langle A(t_{1})\mathcal{B}(t_{2})a_{k_{2}}^{\dag
}(t_{2})...a_{k_{m}}^{\dag }(t_{2})\right\rangle  \nonumber \\
&&+...+\delta _{k,k_{m}}\left\langle A(t_{1})\mathcal{B}(t_{2})a_{k_{1}}^{%
\dag }(t_{2})a_{k_{2}}^{\dag }(t_{2})...a_{k_{m-1}}^{\dag
}(t_{2})\right\rangle  \nonumber \\
&&+\left\langle A(t_{1})\mathcal{B}(t_{2})a_{k_{1}}^{\dag
}(t_{2})a_{k_{2}}^{\dag }(t_{2})...a_{k_{m}}^{\dag
}(t_{2})a_{ks}(t_{2})\right\rangle .  \label{PL22}
\end{eqnarray}%
Introducing this relation in (\ref{PL22}) \ it is not difficult to observe
that the new expression for correlation $\left\langle
A(t_{1})a_{k}(t)B(t_{2})\right\rangle $ coincides with (\ref{L21}). The
similar procedure of permutation of vacuum part of creation operator $%
a_{k}^{\dag }(t)$ demonstrates the identity (\ref{L22}) of Lemma 2.
\end{proof}

According with lemma (\ref{L21}) we obtain the following expression for
correlation function

\begin{eqnarray}
&&\langle \lbrack D_{j}^{+}(t),O(t)]a_{k}(t)D_{n}^{-}(t-\tau
)a_{k_{1}}^{\dagger }(t-\tau )\rangle =\langle \lbrack
D_{j}^{+}(t),O(t)]a_{ks}(t)D_{n}^{-}(t-\tau )a_{k_{1}}^{\dagger }(t-\tau
)\rangle  \nonumber \\
&&-\exp [-i\omega _{k}\tau ]\{\langle \lbrack
D_{j}^{+}(t),O(t)][a_{ks}(t-\tau ),D_{n}^{-}(t-\tau )a_{k_{1}}^{\dagger
}(t-\tau )]\rangle -\delta _{k,k_{1}}\langle \lbrack
D_{j}^{+}(t),O(t)]D_{n}^{-}(t-\tau )\rangle \}.  \label{eq:DD}
\end{eqnarray}%
The next step of elimination of operator $a_{k_{1}}^{\dagger }(t-\tau )$
from this expression must be do taking in-to account the lemma (\ref{L1}).
When the first and second order interaction constants have the same small
magnitude $\lambda \sim (\mathbf{d}_{s},\mathbf{g}_{k})\eqsim q(\omega
_{k_{1}},\omega _{k})$, in Born approximation we take in to account only the
last term of expression (\ref{eq:DD}). the interference contribution of
which is proportional to $\lambda ^{3}$. As follows from the representation (%
\ref{eq:as}) and (\ref{eq:O}) the procedure of elimination mast continue.
Indeed introducing again this equation in the right hand cite of equation (%
\ref{eq:as1}) we obtain the following master equation for arbitrary operator
$O(t)$ in thread approximation on the interaction constant $\lambda $%
\begin{eqnarray}
\frac{d\langle O(t)\rangle }{dt} &=&\sum\limits_{k}\sum%
\limits_{l,j=1}^{N_{r}}\frac{(\mathbf{d}_{a},\mathbf{g}_{k})^{2}}{\hbar ^{2}}%
\int\limits_{0}^{t}d\tau \exp [-i\omega _{k}\tau +i(\mathbf{k,r}_{j}-\mathbf{%
r}_{l})]\langle \lbrack R_{j}^{+}(t),O(t)]R_{l}^{-}(t-\tau )\rangle
\nonumber \\
&+&\sum\limits_{k}\sum\limits_{l,l=1}^{N_{s}}\frac{(\mathbf{d}_{b},\mathbf{g}%
_{k})^{2}}{\hbar ^{2}}\int\limits_{0}^{t}d\tau \exp [-i\omega _{k}\tau +i(%
\mathbf{k,r}_{j}-\mathbf{r}_{l})]\langle \lbrack
S_{j}^{+}(t),O(t)]S_{l}^{-}(t-\tau )\rangle  \nonumber \\
&+&\sum\limits_{k_{1},k_{2}}\sum\limits_{l,j=1}^{N}\frac{(\mathbf{n}_{eg},%
\mathbf{e}_{\lambda _{1}})^{2}(\mathbf{n}_{ei},\mathbf{e}_{\lambda
_{2}})^{2}q^{2}(\omega _{1},\omega _{2})}{\hbar ^{2}}\int\limits_{0}^{t}d%
\tau \langle \lbrack D_{j}^{+}(t),O(t)]D_{l}^{-}(t-\tau )\rangle  \nonumber
\\
&\times &\exp [-i(2\omega _{0}-\omega _{k_{1}}-\omega _{k_{2}})\tau ]\exp [i(%
\mathbf{k}_{1}+\mathbf{k}_{2},\mathbf{r}_{j}-\mathbf{r}_{l})]  \nonumber \\
&+&i\sum\limits_{k,k^{\prime
}}\sum\limits_{n=1}^{N}\sum\limits_{j=1}\sum\limits_{l=1}\frac{(\mathbf{n}%
_{eg},\mathbf{e}_{\lambda }(\mathbf{k}))(\mathbf{n}_{ei},\mathbf{e}_{\lambda
}(\mathbf{k}^{\prime }))q(\omega _{k},\omega _{k^{\prime }})}{\hbar ^{3}}%
\int\limits_{0}^{t}d\tau \int\limits_{0}^{t}d\tau ^{\prime }\exp [i(\mathbf{%
k,r}_{n}-\mathbf{r}_{l})-i\omega _{k}\tau ]  \nonumber \\
&\times &\exp [i(\mathbf{k}^{\prime }\mathbf{,r}_{n}-\mathbf{r}_{l})-i\omega
_{k^{\prime }}\tau ^{\prime }][(\mathbf{d}_{s},\mathbf{g}_{k^{\prime }})(%
\mathbf{d}_{r},\mathbf{g}_{k})\langle \lbrack
D_{n}^{+}(t),O(t)]R_{j}^{-}(t-\tau )S_{l}^{-}(t-\tau ^{\prime })\rangle
\nonumber \\
&+&(\mathbf{d}_{r},\mathbf{g}_{k^{\prime }})(\mathbf{d}_{s},\mathbf{g}%
_{k})\langle \lbrack D_{n}^{+}(t),O(t)]S_{j}^{-}(t-\tau )R_{l}^{-}(t-\tau
^{\prime })\rangle ]  \nonumber \\
&-&i\sum\limits_{k,k^{\prime
}}\sum\limits_{m=1}^{N}\sum\limits_{j=1}\sum\limits_{l=1}\frac{(\mathbf{n}%
_{eg},\mathbf{e}_{\lambda }(\mathbf{k}))(\mathbf{n}_{ei},\mathbf{e}_{\lambda
}(\mathbf{k}^{\prime }))q(\omega _{k},\omega _{k^{\prime }})}{\hbar ^{3}}%
\int\limits_{0}^{t}d\tau ^{\prime }\int\limits_{0}^{t}d\tau \exp [i\omega
_{k}\tau -i(\mathbf{k,r}_{m}-\mathbf{r}_{l})]  \nonumber \\
&\times &\exp [-i\omega _{k}\tau ^{\prime }-i\omega _{k^{\prime }}\tau
^{\prime }+i(\mathbf{k}^{\prime }\mathbf{,r}_{j}-\mathbf{r}_{m})][(\mathbf{d}%
_{s},\mathbf{g}_{k})(\mathbf{d}_{r},\mathbf{g}_{k^{\prime }})\langle
S_{l}^{+}(t-\tau )[R_{j}^{+}(t),O(t)]D_{m}^{-}(t-\tau ^{\prime })\rangle
\nonumber \\
&+&(\mathbf{d}_{s},\mathbf{g}_{k})(\mathbf{d}_{r},\mathbf{g}_{k^{\prime
}})\langle R_{l}^{+}(t-\tau )[S_{j}^{+}(t),O(t)]D_{m}^{-}(t-\tau ^{\prime
})\rangle ]+H.c.(O^{+}\rightarrow O).  \label{eq:O2}
\end{eqnarray}

The traditional Born-Marcov approximation in the right hand site of equation
( \ref{eq:O2}) give us the divergent functions. In order to understood this
we approximate the right hand site of equation (\ref{eq:as}) with following
expression

\begin{eqnarray}
a_{ks}(t) &=&\frac{(\mathbf{d}_{a},\mathbf{g}_{k})}{\hbar }%
\sum\limits_{l=1}^{N_{a}}R_{l}^{-}(t)\exp [-i(\mathbf{k},\mathbf{r}%
_{l})\zeta ^{\ast }(\omega _{k}-\omega _{a})  \nonumber \\
&&+\frac{(\mathbf{d}_{b},\mathbf{g}_{k})}{\hbar }\sum%
\limits_{j=1}^{N_{b}}S_{j}^{-}(t)\exp [-i(\mathbf{k},\mathbf{r}_{j})\zeta
^{\ast }(\omega _{k}-\omega _{b})  \nonumber \\
&&+2\sum\limits_{n=1}^{N_{b}}\sum\limits_{k_{1}}\frac{(\mathbf{n}_{eg},%
\mathbf{e}_{\lambda _{1}})(\mathbf{n}_{ei},\mathbf{e}_{\lambda })q(\omega
_{1},\omega )}{\hbar }  \nonumber \\
&&\times D_{n}^{-}(t)a_{k_{1}}^{\dagger }(t)\exp [-i(\mathbf{k}_{1}+\mathbf{k%
},\mathbf{r}_{n})\zeta ^{\ast }(\omega _{k}+\omega _{k_{1}}-2\omega _{0}),
\label{eq:as1}
\end{eqnarray}%
in the Born-Marcovian approximation \cite{H_1954} , \cite{GH_1982}. The
small parameter in this approximation is the ratio of retardation time to
cooperative spontaneous emission times of the subsystem, $\tau /\tau _{i}<<1$%
. Here $i\zeta (x)=iP/x+\pi \delta (x)$ is the Heitler function \cite%
{GH_1982},\cite{AEI_1992}. represents k-summation in analogy with Cauchy
principal value \cite{H_1954}. Introducing the operators (\ref{eq:as1}) in
equation (\ref{eq:O}) and eliminating the boson operators of EMF, it is
obtain the following equation for operator $O(t)$ in Born-Marcov
approximation \

\begin{eqnarray}
\frac{d}{dt}\langle O(t)\rangle &=&\sum\limits_{k}\sum\limits_{l,j=1}^{N_{r}}%
\frac{(\mathbf{d}_{a},\mathbf{g}_{k})^{2}}{\hbar ^{2}}\langle \lbrack
R_{j}^{+}(t),O(t)]R_{l}^{-}(t)\rangle  \nonumber \\
&\times &\exp [i(\mathbf{k,r}_{j}-\mathbf{r}_{l})]i\zeta ^{\ast }(\omega
_{r}-\omega _{k})+\sum\limits_{k}\sum\limits_{l,j=1}^{N_{s}}\frac{(\mathbf{d}%
_{a},\mathbf{g}_{k})(\mathbf{d}_{b},\mathbf{g}_{k})}{\hbar ^{2}}  \nonumber
\\
&\times &\langle \lbrack S_{j}^{+}(t),O(t)]S_{l}^{-}(t)\rangle \exp [i(%
\mathbf{k,r}_{j}-\mathbf{r}_{l})]i\zeta ^{\ast }(\omega _{s}-\omega _{k})
\nonumber \\
&+&\sum\limits_{k,k^{\prime }}\sum\limits_{l,j=1}^{N}\frac{(\mathbf{n}_{eg},%
\mathbf{e}_{\lambda })^{2}(\mathbf{n}_{ei},\mathbf{e}_{\lambda ^{\prime
}})^{2}q^{2}(\omega _{k},\omega _{k^{\prime }})}{\hbar ^{2}}  \nonumber \\
&\times &\langle \lbrack D_{j}^{+}(t),O(t)]D_{l}^{-}(t)\rangle \exp [i(%
\mathbf{k-k}^{\prime }\mathbf{,r}_{j}-\mathbf{r}_{l})]i\zeta ^{\ast }(\omega
_{a}-\omega _{k}-\omega _{k^{\prime }})  \nonumber \\
&+&2i\sum\limits_{k,k^{\prime
}}\sum\limits_{n=1}^{N}\sum\limits_{l=1}^{N_{s}}\sum\limits_{l=1}^{N_{r}}%
\frac{(\mathbf{d}_{a},\mathbf{g}_{k^{\prime }})(\mathbf{d}_{b},\mathbf{g}%
_{k})(\mathbf{n}_{eg},\mathbf{e}_{\lambda })(\mathbf{n}_{ei},\mathbf{e}%
_{\lambda ^{\prime }})q(\omega _{k},\omega _{k^{\prime }})}{\hbar ^{3}}
\nonumber \\
&\times &\langle \lbrack D_{n}^{+}(t),O(t)]R_{j}^{-}(t)S_{l}^{-}(t)\rangle
\nonumber \\
&\times &\exp [i(\mathbf{k,r}_{n}-\mathbf{r}_{l})+i(\mathbf{k}^{\prime }%
\mathbf{,r}_{n}-\mathbf{r}_{l})]i\zeta ^{\ast }(\omega _{r}-\omega
_{k})i\zeta ^{\ast }(\omega _{s}-\omega _{k^{\prime }})  \nonumber \\
&-&i\sum\limits_{k,k^{\prime
}}\sum\limits_{m=1}^{N}\sum\limits_{l=1}\sum\limits_{j=1}\frac{(\mathbf{d}%
_{a},\mathbf{g}_{k_{1}})(\mathbf{d}_{b},\mathbf{g}_{k_{2}})(\mathbf{n}_{eg},%
\mathbf{e}_{\lambda _{1}})(\mathbf{n}_{ei},\mathbf{e}_{\lambda
_{2}})q(\omega _{1},\omega _{2})}{\hbar ^{3}}  \nonumber \\
&\times &[\langle S_{l}^{+}(t)[R_{j}^{+}(t),O(t)]D_{m}^{-}(t)\rangle i\zeta
^{\ast }(\omega _{r}-\omega _{k})i\zeta (\omega _{s}-\omega _{k^{\prime }})
\nonumber \\
&+&\langle R_{l}^{+}(t)[S_{j}^{+}(t),O(t)]D_{m}^{-}(t)\rangle i\zeta ^{\ast
}(\omega _{s}-\omega _{k})i\zeta (\omega _{r}-\omega _{k^{\prime }})]
\nonumber \\
\times &&\exp [i(\mathbf{k,r}_{j})+i(\mathbf{k}^{\prime }\mathbf{,r}_{l})-i(%
\mathbf{k+k}^{\prime },\mathbf{r}_{m})]+H.C.(O^{+}\rightarrow O).
\label{eq:O3}
\end{eqnarray}

In the right hand part of the equation (\ref{eq:O3}) the third order terms
contain the resonances between the single photon radiators $A$ ,$B$ and
two-photon radiator $D$ described by the correlation functions $\langle
S_{l}^{+}(t)[R_{j}^{+}(t),O(t)]D_{m}^{-}(t)\rangle $, $\langle \lbrack
D_{n}^{+}(t),O(t)]R_{j}^{-}(t)S_{l}^{-}(t)\rangle $ and $\langle
R_{l}^{+}(t)[S_{j}^{+}(t),O(t)]D_{m}^{-}(t)\rangle $. As it is observed from
equation (\ref{eq:O3}), these terms contain the product of the functions $%
[P/(\omega _{b}-\omega _{k})][P/(\omega _{b}-\omega _{k^{\prime }})]$ which
describe the principal value in the integration procedure on the variables $%
k $ and $k^{\prime }$. It is not difficult to observe that these integrals
become divergent expressions In order to avoid these divergence in Appendix1
it is proposed the integration procedure which takes in to account the
retardation between the radiators in the representation of right hand site
of equation (\ref{eq:O2}). According with Appendix1 the right hand site of
master equation (\ref{eq:O2}) takes the following non-divergent form

\begin{eqnarray}
\frac{d}{dt}\langle O(t)\rangle &=&\frac{1}{2\tau _{r}}\sum%
\limits_{l,j=1}^{N_{r}}\chi _{r}(j,l)\langle \lbrack
R_{j}^{+}(t),O(t)]R_{l}^{-}(t)\rangle +\frac{1}{2\tau _{s}}%
\sum\limits_{l,l=0}^{N_{s}}\chi _{s}(j,l)\langle \lbrack
S_{j}^{+}(t),O(t)]S_{l}^{-}(t)\rangle  \nonumber \\
&+&\frac{1}{2\tau _{b}}\sum\limits_{l,j=1}^{N}\chi _{d}(j,l)\langle \lbrack
D_{j}^{+}(t),O(t)]D_{l}^{-}(t)\rangle  \nonumber \\
&+&\frac{i}{2\tau _{bsr}}\sum\limits_{m=1}^{N}\sum\limits_{l=1}^{N_{r}}\sum%
\limits_{l=0}^{N_{s}}U(j,l,m)\langle \lbrack
D_{m}^{+}(t),O(t)]R_{j}^{-}(t)S_{l}^{-}(t)\rangle  \nonumber \\
&-&\frac{i}{4\tau _{sbr}}\sum\limits_{m=1}^{N}\sum\limits_{j=1}^{N_{r}}\sum%
\limits_{l=0}^{N_{s}}V(j,l.m)[\langle
S_{l}^{+}(t)[R_{j}^{+}(t),O(t)]D_{m}^{-}(t)\rangle +\langle
R_{j}^{+}(t)[S_{l}^{+}(t),O(t)]D_{m}^{-}(t)\rangle ]  \nonumber \\
&+&H.C.(O^{+}\rightarrow O).  \label{eq:O4}
\end{eqnarray}

Here the spontaneous emission $\tau _{i}$ and exchange integral between
radiators $j$ and $l\ $, $\chi _{i}(j,l)$ are defined in expressions (\ref%
{eq:exp1}), (\ref{eq:exp2}), (\ref{eq:U}) and (\ref{eq:Vjlm}) of Appendix.
This equation can be used for description of interaction between the dipole
forbidden and dipole active systems of radiators.

\section{Kinetic equations for correlation functions}

\begin{figure}[t]
\begin{center}
\includegraphics{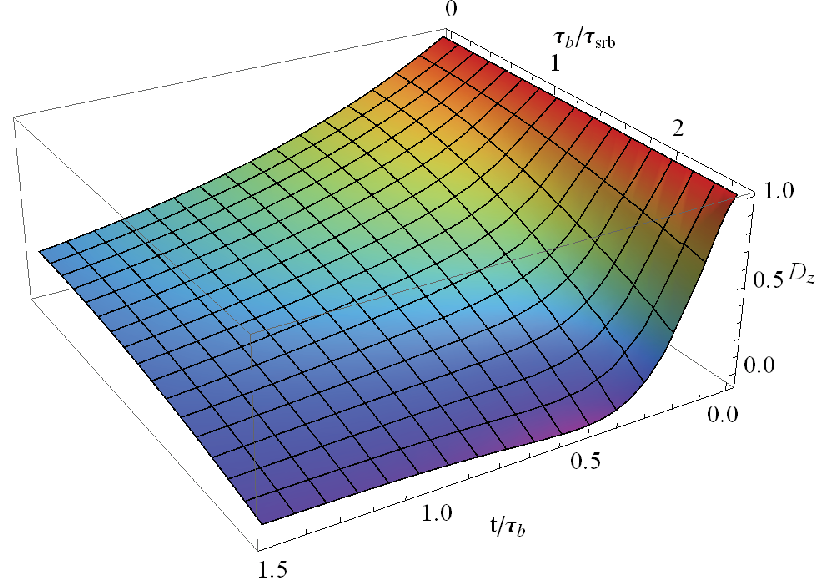}
\end{center}
\caption{The decay law of inversion, $D_{z}(t)$, of dipole forbidden
radiator stimulated by single photon decay processes of two radiators for
following parameters of the system: relative decay rates of S- and R- atoms $%
\protect\tau_{b}/\protect\tau_{s}=\protect\tau_{b}/\protect\tau_{r}=5$.}
\label{figure_2a}
\end{figure}

In order to found the correlation process between dipole forbidden
transitions of $D$ subsystem and dipole active transitions in $S$ and $R$
subsystems of radiators let us found the equations for arbitrary atomic
correlation functions of subsystems of radiators. According with the
generalized equation (\ref{eq:O4}) it is obtain the following chain of
equation for atomic correlations
\begin{eqnarray}
\frac{d}{dt}\langle R_{zj}(t)\rangle &=&-\frac{1}{2\tau _{r}}%
\sum\limits_{l,=1}^{N_{a}}[\chi _{r}(j,l)\langle
R_{j}^{+}(t)R_{l}^{-}(t)\rangle +\chi _{r}^{\ast }(j,l)\langle
R_{l}^{+}(t)R_{j}^{-}(t)\rangle ]  \nonumber \\
&&+\frac{i}{4\tau _{sbr}}\sum\limits_{m=1}^{N}\sum%
\limits_{l=0}^{N_{s}}[V(j,l.m)\langle
S_{l}^{+}(t)R_{j}^{+}(t)D_{m}^{-}(t)\rangle -V^{\ast }(j,l.m)\langle
D_{m}^{+}(t)R_{j}^{-}(t)S_{l}^{-}(t)\rangle ],  \label{eq:Ch1}
\end{eqnarray}%
\begin{eqnarray}
\frac{d}{dt}\langle S_{zl}(t)\rangle &=&-\frac{1}{2\tau _{s}}%
\sum\limits_{p=0}^{N_{s}}\chi _{s}(l,p)[\langle
S_{l}^{+}(t)S_{p}^{-}(t)\rangle +\chi _{s}^{\ast }(l,p)\langle
S_{p}^{+}(t)S_{l}^{-}(t)\rangle ]  \nonumber \\
&+&\frac{i}{4\tau _{sbr}}\sum\limits_{m=1}^{N}\sum%
\limits_{l=0}^{N_{s}}[V(j,l.m)\langle
S_{l}^{+}(t)R_{j}^{+}(t)D_{m}^{-}(t)\rangle -V^{\ast }(j,l.m)\langle
D_{m}^{+}(t)R_{j}^{-}(t)S_{l}^{-}(t)\rangle ],  \label{eq:Ch2}
\end{eqnarray}%
\begin{eqnarray}
\frac{d}{dt}\langle D_{zn}(t)\rangle &=&-\frac{1}{2\tau _{b}}%
\sum\limits_{l=1}^{N}[I_{^{b}}(j,l)\langle D_{n}^{+}(t)D_{l}^{-}(t)\rangle
+I_{^{b}}^{\ast }(j,l)\langle D_{l}^{+}(t)D_{n}^{-}(t)\rangle  \nonumber \\
&-&\frac{i}{2\tau _{srb}}\sum\limits_{j=1}^{N_{r}}\sum%
\limits_{l=0}^{N_{s}}[U(j,l,n)\langle
D_{n}^{+}(t)R_{j}^{-}(t)S_{l}^{-}(t)\rangle -U^{\ast }(j,l,n)\langle
S_{l}^{+}(t)R_{j}^{+}(t)D_{n}^{-}(t)\rangle ,  \label{eq:Ch3}
\end{eqnarray}

\begin{eqnarray}
\frac{d}{dt}\langle R_{j}^{+}(t)R_{l}^{-}(t)\rangle &=&\frac{1}{\tau _{r}}%
\sum\limits_{n=1}^{N_{r}}[\chi _{r}(l,n)\langle
R_{j}^{+}(t)R_{zl}(t)R_{n}^{-}(t)\rangle +\chi _{r}(n,j)\langle
R_{n}^{+}(t)R_{zj}(t)R_{l}^{-}(t)\rangle ]  \nonumber \\
&-&\frac{1}{2\tau _{sbr}}\sum\limits_{m=1}^{N}\sum%
\limits_{k=0}^{N_{s}}[V(j,k,m)\langle
S_{k}^{+}(t)R_{j}^{+}(t)R_{zl}(t)D_{m}^{-}(t)\rangle  \nonumber \\
&+&V^{\ast }(j,k.m)\langle
D_{m}^{+}(t)R_{zl}(t)R_{j}^{-}(t)S_{k}^{-}(t)\rangle ];  \label{eq:Ch4}
\end{eqnarray}

\begin{eqnarray}
\frac{d}{dt}\langle S_{j}^{+}(t)S_{l}^{-}(t)\rangle &=&\frac{1}{\tau _{r}}%
\sum\limits_{n=1}^{N_{s}}[\chi _{r}(l,n)\langle
S_{j}^{+}(t)S_{zl}(t)S_{n}^{-}(t)\rangle +\chi _{r}(n,j)\langle
S_{n}^{+}(t)S_{zj}(t)S_{l}^{-}(t)\rangle ]  \nonumber \\
&-&\frac{i}{2\tau _{sbr}}\sum\limits_{m=1}^{N}\sum%
\limits_{k=1}^{N_{r}}[V(k,l.m)\langle
R_{k}^{+}(t)S_{j}^{+}(t)S_{zl}(t)D_{m}^{-}(t)\rangle  \nonumber \\
&-&V^{\ast }(k,j.m)\langle
D_{m}^{+}(t)S_{zj}(t)S_{l}^{-}(t)R_{k}^{-}(t)\rangle ];  \label{eq:Ch5}
\end{eqnarray}%
\begin{eqnarray}
\frac{d}{dt}\langle D_{n}^{+}(t)D_{l}^{-}(t)\rangle &=&\frac{1}{\tau _{b}}%
\sum\limits_{m=1}^{N}[I_{^{b}}(l,m)\langle
D_{n}^{+}(t)D_{zl}(t)D_{m}^{-}(t)\rangle +I_{^{b}}^{\ast }(n,m)\langle
D_{m}^{+}(t)D_{zn}(t)D_{l}^{-}(t)\rangle ]  \nonumber \\
&+&\frac{i}{\tau _{bsr}}\sum\limits_{j=1}^{N_{r}}\sum%
\limits_{k=0}^{N_{s}}[U(j,k,m)\langle
D_{n}^{+}(t)D_{zl}(t)R_{j}^{-}(t)S_{k}^{-}(t)\rangle  \nonumber \\
&+&U^{\ast }(j,k,m)\langle
S_{k}^{+}(t)R_{j}^{+}(t)D_{zn}(t)D_{l}^{+}(t)\rangle ].  \label{eq:Ch6}
\end{eqnarray}

\begin{eqnarray}
\frac{d}{dt}i\langle D_{m}^{+}(t)S_{l}^{-}(t)R_{k}^{-}(t)\rangle &=&\frac{1}{%
\tau _{bsr}}\sum\limits_{j=1}^{N_{r}}\sum\limits_{n=1}^{N_{s}}[U^{\ast
}(j,n,m)\langle
S_{n}^{+}(t)R_{j}^{+}(t)D_{zm}(t)S_{l}^{-}(t)R_{k}^{-}(t)\rangle  \nonumber
\\
&+&\frac{1}{2\tau _{sbr}}\sum\limits_{n=1}^{N}\sum\limits_{j=1}[V(j,l.n)%
\langle S_{zl}(t)R_{j}^{+}(t)R_{k}^{-}(t)D_{m}^{+}(t)D_{n}^{-}(t)\rangle
\nonumber \\
&+&V(k,j.n)\langle
D_{m}^{+}(t)D_{n}^{-}(t)R_{zk}(t)S_{j}^{+}(t)S_{l}^{-}(t)\rangle ]  \nonumber
\\
&+&i\sum\limits_{j=1}\bigl[\frac{1}{\tau _{s}}\chi _{r}(j,l)\langle
D_{m}^{+}(t)S_{zl}(t)S_{j}^{-}(t)R_{k}^{-}(t)\rangle  \nonumber \\
&&+\frac{1}{\tau _{r}}\chi _{r}(j,k)\langle
D_{m}^{+}(t)R_{zk}(t)S_{l}^{-}(t)R_{j}^{-}(t)\rangle  \nonumber \\
&&+\frac{1}{\tau _{b}}I_{^{b}}^{\ast }(m,l)\langle
D_{j}^{+}(t)D_{zm}(t)S_{j}^{-}(t)R_{k}^{-}(t)\rangle \bigr]  \label{eq:Ch7}
\end{eqnarray}%
Let us consider the interaction three different atoms in interaction through
vacuum EMF. Introducing the exited numbers for the \ atomic subsystems $%
\langle N_{\alpha }\rangle =\langle J_{z\alpha }(t)\rangle +0.5$ (here $%
J\leftrightarrow S,\ R,\ D$ $\ \alpha =s,\ r,\ d$ ) and correlation function
between the atoms $\langle F\rangle =i[\langle
D^{+}(t)S^{-}(t)R^{-}(t)\rangle -\langle S^{+}(t)R^{+}(t)D^{-}(t)\rangle ]$\
\ we can obtain the closed system of equations from the chain of equations (%
\ref{eq:Ch1}-\ref{eq:Ch7}). Indeed considering that the distance between the
radiators is smaller then radiation wavelength: $\Re $ $\{U(j,k,m)\}=\Re
\{V(j,k,m)\}=1$ , we obtain the following closed system of equation

\begin{eqnarray*}
\frac{d}{dt}\langle N_{s}(t)\rangle &=&-\frac{\langle N_{s}\rangle }{\tau
_{s}}-\frac{1}{2\tau _{sbr}}\langle F\rangle ; \\
\frac{d}{dt}\langle N_{r}(t)\rangle &=&-\frac{\langle N_{r}\rangle }{\tau
_{r}}-\frac{1}{2\tau _{sbr}}\langle F\rangle ; \\
\frac{d}{dt}\langle N_{d}(t)\rangle &=&-\frac{\langle N_{d}\rangle }{\tau
_{d}}-\frac{1}{\tau _{sbr}}\langle F\rangle ; \\
\frac{d}{dt}\langle F(t)\rangle &=&-\bigl[\frac{1}{2\tau _{s}}+\frac{1}{%
2\tau _{r}}+\frac{1}{2\tau _{d}}\bigr]\langle F(t)\rangle \\
&+&\frac{1}{\tau _{bsr}}[6\langle N_{s}N_{r}N_{d}\rangle -2\langle
N_{s}N_{r}\rangle -\langle N_{s}N_{d}\rangle -\langle N_{r}N_{d}\rangle ], \\
\frac{d}{dt}\langle N_{s}N_{r}N_{d}\rangle &=&-[\frac{1}{\tau _{r}}+\frac{1}{%
\tau _{b}}+\frac{1}{\tau _{s}}]\langle N_{s}N_{r}N_{d}\rangle , \\
\frac{d}{dt}\langle N_{s}N_{r}\rangle &=&-[\frac{1}{\tau _{r}}+\frac{1}{\tau
_{s}}]\langle N_{s}N_{r}\rangle , \\
\frac{d}{dt}\langle N_{s}N_{d}\rangle &=&-[\frac{1}{\tau _{r}}+\frac{1}{\tau
_{b}}]\langle N_{s}N_{b}\rangle , \\
\frac{d}{dt}\langle N_{s}N_{d}\rangle &=&-[\frac{1}{\tau _{r}}+\frac{1}{\tau
_{b}}]\langle N_{s}N_{b}\rangle ,
\end{eqnarray*}%
in which the new correlation functions between the atomic excitation is
introduced $\langle \hat{N}_{s}\hat{N}_{r}\hat{N}_{d}\rangle $, $\langle
\hat{N}_{s}\hat{N}_{r}\rangle $, $\langle \hat{N}_{s}\hat{N}_{d}\rangle $,
and $\langle \hat{N}_{r}\hat{N}_{d}\rangle $. This system of equation is
exactly solvable. The solution is

\begin{eqnarray*}
\langle N_{s}N_{r}N_{d}\rangle &=&\exp (-At);\ \ \ \langle N_{s}N_{r}\rangle
=\exp (-Bt),\langle N_{s}N_{d}\rangle =\exp (-Ct);\ \ \ \langle
N_{d}N_{r}\rangle =\exp (-Dt) \\
\langle N_{i}N_{j}\rangle &=&\exp \bigl[-\bigl(\frac{1}{\tau _{i}}+\frac{1}{%
\tau _{j}}\bigr)t\bigr],\ \ \ i,j\rightarrow s,\ r,\ b; \\
\langle F(t)\rangle &=&\frac{\exp (-At/2)}{\tau _{bsr}}\bigl[6\frac{1-\exp
[-At/2]}{A}-4\frac{1-\exp [-(B-\tau _{b}^{-1})t/2]}{(B-\tau _{b}^{-1})} \\
&&-\frac{2(1-\exp [-(C-\tau _{r}^{-1})t/2])}{(C-\tau _{r}^{-1})}-\frac{%
2(1-\exp [-(C-\tau _{s}^{-1})t/2])}{(C-\tau _{s}^{-1})}\bigr]; \\
\langle N_{i}\rangle &=&\exp [-t/\tau _{i}]-\frac{1}{\tau _{bsr}}%
\int\limits_{0}^{t}\exp [-(t-t^{\prime })/\tau _{i}]\langle F(t^{\prime
})\rangle ,\ \ \ \ \ i\equiv s,\ r,\ d,
\end{eqnarray*}%
where the collective rates are defined $A=1/\tau _{r}+1/\tau _{b}+1/\tau
_{s} $; \ $B=1/\tau _{r}+1/\tau _{s}$;\ $C=1/\tau _{b}+1/\tau _{s}$;\ \ $%
D=1/\tau _{b}+1/\tau _{s}$. The solution of this system of equation is
plotted in figure \ref{figure_2a}. It is observed the influence of single photon
transition on the two quanta transitions. This influence drastically depends
on the cooperative rate $1/\tau _{bsr}$.

\begin{figure}[t]
\begin{center}
\includegraphics{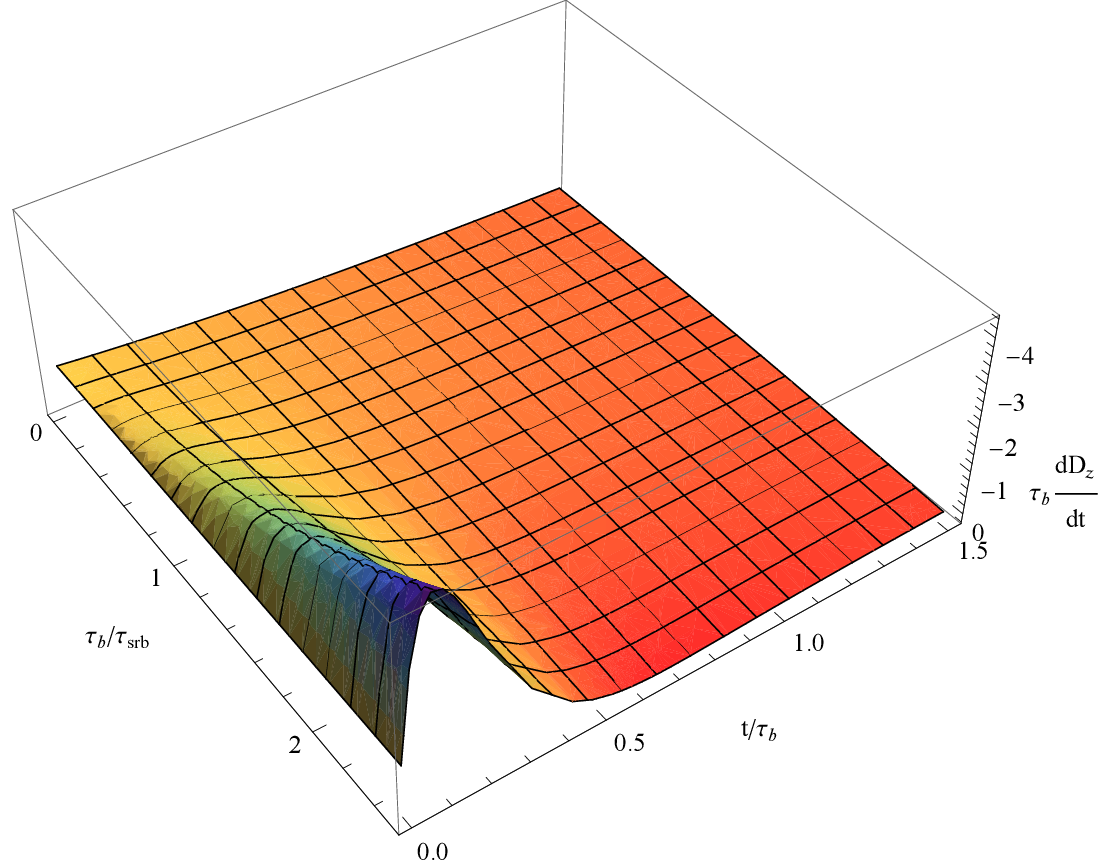}
\end{center}
\caption{The two-photon decay rate $dD_{z}(t)/dt$ stimulated by single
photon processes for same parameter of the system as in figure \protect\ref%
{figure_2a}.}
\label{figure_2b}
\end{figure}

As follows from figure \ref{figure_2b} the cooperative exchanges between the
radiator accelerate the two-photon decay processes so that the the
derivation $-\frac{dD_z}{dt}$ achieved the maximal value decreasing after
that till zero value.

Neglecting the quantum fluctuations of inversion operators $\langle
R_{z_{j}}\rangle $, $\langle S_{zl}\rangle ,$ $\langle D_{zn}\rangle $,
dipole-dipole correlations\ between the same radiators $\langle
R_{l}^{+}(t)R_{j}^{-}(t)\rangle $, $\langle S_{l}^{+}(t)S_{p}^{-}(t)\rangle $
and $\langle D_{n}^{+}(t)D_{l}^{-}(t)\rangle $ and between subsystems $%
\langle S_{l}^{+}(t)R_{j}^{+}(t)D_{m}^{-}(t)\rangle $, $\langle
D_{m}^{+}(t)R_{j}^{-}(t)S_{l}^{-}(t)\rangle $ we can de-correlated the chain
of equations (\ref{eq:Ch1}-\ref{eq:Ch7}) in order to obtain the closed
system of equations
\begin{eqnarray}
for\ \ \ J\leftrightarrow S,\ R,\ D;\ \langle
J_{j}^{+}(t)J_{zl}(t)J_{m}^{-}(t)\rangle &=&\langle J_{zl}(t)\rangle \langle
J_{j}^{+}(t)J_{m}^{-}(t)\rangle \ \ \ j\neq l\neq m;  \nonumber \\
\langle S_{k}^{+}(t)R_{j}^{+}(t)R_{zl}(t)D_{m}^{-}(t)\rangle &=&\langle
R_{zl}(t)\rangle \langle S_{k}^{+}(t)R_{j}^{+}(t)D_{m}^{-}(t)\rangle _{l\neq
j}-\delta _{l,j}\langle S_{k}^{+}(t)R_{j}^{+}(t)D_{m}^{-}(t)\rangle ;
\nonumber \\
\langle D_{n}^{+}(t)D_{zl}(t)R_{j}^{-}(t)S_{k}^{-}(t)\rangle &=&\langle
D_{zl}(t)\rangle \langle D_{n}^{+}(t)R_{j}^{-}(t)S_{k}^{-}(t)\rangle _{l\neq
n}-\delta _{l,n}\langle D_{n}^{+}(t)R_{j}^{-}(t)S_{k}^{-}(t)\rangle ;
\nonumber \\
\langle S_{l}^{+}(t)R_{j}^{+}(t)D_{zn}(t)R_{p}^{-}(t)S_{k}^{-}(t)\rangle
&=&\langle D_{zn}(t)\rangle \langle R_{j}^{+}(t)R_{p}^{-}(t)\rangle \langle
S_{l}^{+}(t)S_{k}^{-}(t)\rangle _{l\neq k;j\neq p}  \nonumber \\
&&+\delta _{j,p}\delta _{l,k}\langle
D_{zn}(t)(R_{zj}(t)+0.5)(S_{zl}(t)+0.5)\rangle ;  \nonumber \\
\langle D_{m}^{+}(t)D_{n}^{-}(t)S_{zl}(t)R_{j}^{+}(t)R_{k}^{-}(t)\rangle
&=&\langle S_{zl}(t)\rangle \langle D_{m}^{+}(t)D_{n}^{-}(t)\rangle \langle
R_{j}^{+}(t)R_{k}^{-}(t)\rangle _{m\neq n;j\neq k}  \nonumber \\
&&+\delta _{m,n}\delta _{j,k}\langle
(D_{zn}(t)+0.5)S_{zl}(t)(R_{zj}(t)+0.5)\rangle ;  \nonumber \\
\langle D_{m}^{+}(t)D_{n}^{-}(t)R_{zj}(t)S_{l}^{+}(t)S_{k}^{-}(t)\rangle
&=&\langle R_{zj}(t)\rangle \langle D_{m}^{+}(t)D_{n}^{-}(t)\rangle \langle
S_{l}^{+}(t)S_{k}^{-}(t)\rangle _{m\neq n;l\neq k}  \nonumber \\
&&\delta _{m,n}\delta _{l,k}\langle
(D_{zn}(t)+0.5)R_{zj}(t)(S_{zl}(t)+0.5)\rangle .  \label{eq:DCM}
\end{eqnarray}

Taking in to account the de-correlation (\ref{eq:DCM}), we obtain the
following closed system of equations%
\begin{eqnarray*}
\frac{d}{dt}R_{z}(t) &=&-\frac{1}{\tau _{r}}%
\{N_{r}(N_{r}+2)/4-R_{z}^{2}+R_{z}\}-\frac{1}{2\tau _{sbr}}F, \\
\frac{d}{dt}S_{z}(t) &=&-\frac{1}{\tau _{r}}%
\{N_{s}(N_{s}+2)/4-S_{z}^{2}+S_{z}\}-\frac{1}{2\tau _{sbr}}F, \\
\frac{d}{dt}D_{z}(t) &=&-\frac{1}{\tau _{r}}\{N(N+2)/4-D_{z}^{2}+D_{z}\}-%
\frac{1}{\tau _{sbr}}F; \\
\frac{d}{dt}F &=&\bigl[\frac{1}{\tau _{s}}[S_{z}(t)-1]+\frac{1}{\tau _{r}}%
[R_{z}(t)-1]+\frac{1}{\tau _{d}}[D_{z}(t)-1]\bigr]F \\
&+&\frac{1}{\tau _{bsr}}\bigl[2D_{z}\{N_{s}^{2}/4-S_{z}^{2}\}%
\{N_{r}^{2}/4-R_{z}^{2}\} \\
&+&R_{z}\{N_{s}^{2}/4-S_{z}^{2}\}\{N^{2}/4-D_{z}^{2}\}+S_{z}%
\{N_{r}^{2}/4-R_{z}^{2}\}\{N^{2}/4-D_{z}^{2}\} \\
&+&(4N_{s}N_{r}N\exp [-A\ast t]-N_{s}\ast N_{r}\ast \exp[-Bt]-0,5N_{s}N\exp
[-Ct] \\
&-&0.5N_{r}N\exp [-Dt]\bigr].
\end{eqnarray*}

\begin{figure}[t]
\begin{center}
\includegraphics{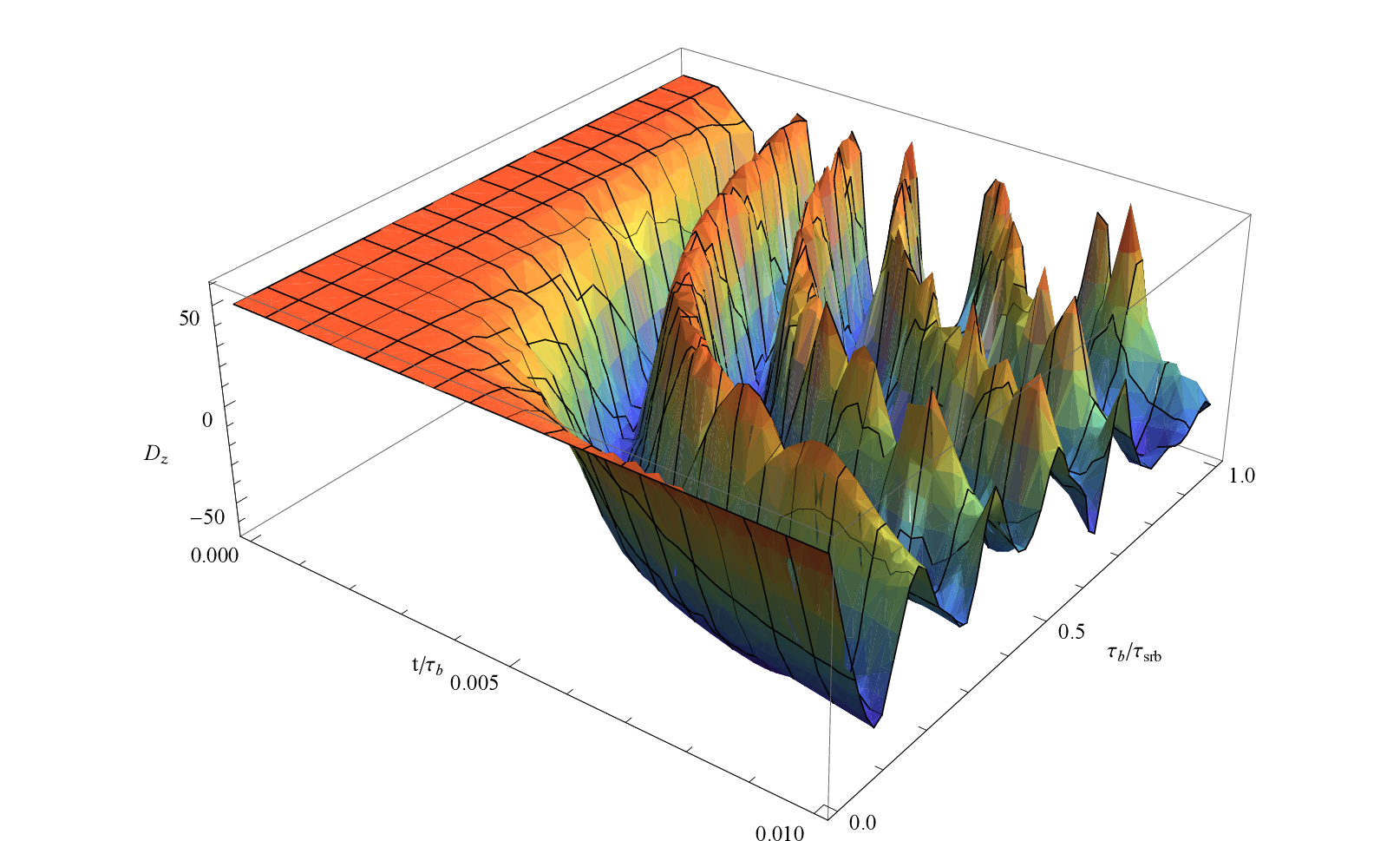}
\end{center}
\caption{The time dependence of inversion $D_{z}(t)$ for dipole forbidden
radiator subsystem for following values of the parameter of the system:nuber
of atoms in the subsystems $S,R$ and $D$ are, $N_{s}=N_{r}=N=50$
respectively; the relative decay times of the subsystems are $\protect\tau %
_{b}/\protect\tau _{s}=\protect\tau _{b}/\protect\tau _{r}=6$ ; the coupling
parameter of these three system is changed between $0$ and $1$. The
oscillatory behavior of decay rate is observed.}
\label{figure_3a}
\end{figure}

\begin{figure}[t]
\begin{center}
\includegraphics{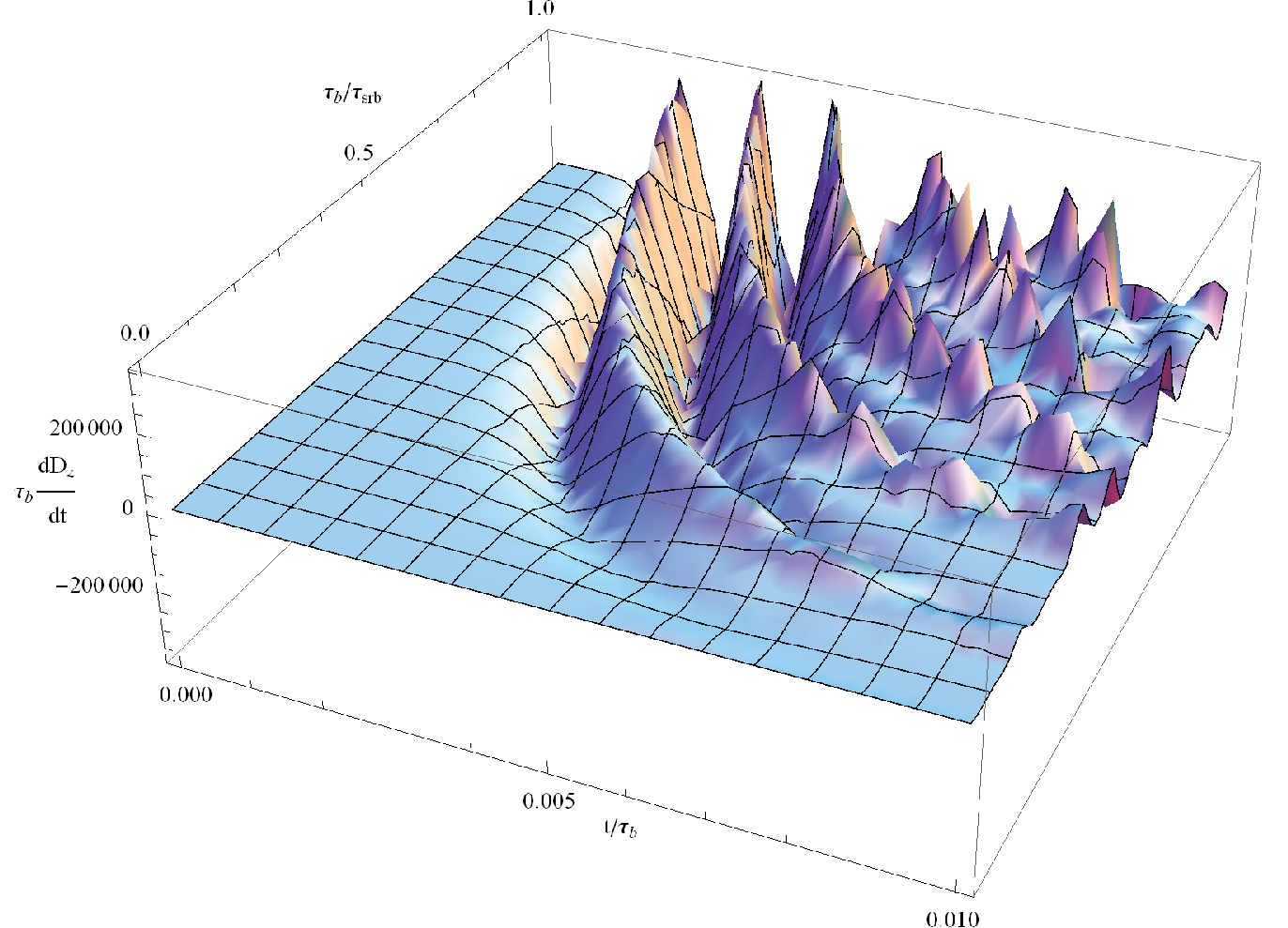}
\end{center}
\caption{The cooperative decay rate of inversion $\frac{dD_{z}(t)}{dt}$ of
dipole forbidden radiator subsystem for same values of the parameters of the
system as in figure \protect\ref{figure_3a}. The increasing of decay rate of
bi-photons is observed.}
\label{figure_3b}
\end{figure}

From this system of equations follows the oscillatory behavior of the decay
rate of the inversion $D_{z}$. Taking in to account the following relative
expressions of the decay rates we obtain the numerical simulation of the
inversion $D_{z}$ and its derivative (see figure \ref{figure_3a} and figure \ref{figure_3b}%
). As follows from this system of equation the increasing of decay rate of
two-photon spontaneous emission is possible under the influence of single
photon cooperative emission of two atomic subsystems. In figures \ref%
{figure_3a} is plotted the time dependence of the inversion $<D_{z}(t)>$ of
dipole forbidden radiators as function of the relative coupled parameter
between the radiators $\tau _{b}/\tau _{srb}$.

The same dependence is represented in figure \ref{figure_3b} for the
intensity of two-photon emission proportional to $-d<D_{z}(t)>/dt$. As
follows from these plots it is observed the mutual influences between single
and two-photon super-radiance processes of three particle interaction. This
effect plays an important role in the collective decay process of the
systems of radiators with the dimension smaller than wavelength.

\section{Conclusion}

In this paper the effective interaction between three radiator subsystems in
two-photon resonance is found using the method of elimination of operators
of vacuum field. The new cooperative interaction between dipole-forbidden
atomic subsystem and two-dipole active subsystems of radiators was proposed.
The master equation \ref{eq:O4}, which describes the energy dissipation from
the system due to mutual interaction between the radiators through the
vacuum of electromagnetic field, was obtained. Using the chain of equation%
\ref{eq:Ch1}-\ref{eq:Ch7}, which describes the cooperative interaction
between three radiator subsystems, it is obtained the closed system of
equations for three radiators. Neglecting the quantum fluctuation of the
inversion, the de-correlation method of the this chain of equation is
proposed \ref{eq:DCM} in order to describe numerically the behavior of
mutual influences of single and two-photon super-radiance processes. As a
consequence of effective interaction between three radiators through
two-photon resonance processes of inverted systems increase substantially in
process cooperative decay of the system The three particle exchange integral
has been established and the influence of this effect on the behavior
cooperative decay of the atomic subsystems was estimated (see figure \ref%
{figure_3a} and figure \ref{figure_3b}). Similar experimental situation can
bi realized in exited atomic (for example transitions in Cs atoms \cite%
{BH_1995}) or nuclei (for example $^{193m}Ir$, $^{195m}Pt$ and $^{103m}Rh$
nuclei\cite{CX_2007}) subsystems in resonance interaction through vacuum
field.

\section{ Appendix: Exchange integrals}

In order to estimate all exchange integrals in equation (\ref{eq:O2}) let us
firstly found the well known exchange integral between two radiators in
single photon interaction with vacuum of electromagnetic field. In the first
terms of equation (\ref{eq:O2}) the retardation can be found integration
firstly on the $k$ vector respectively%
\begin{eqnarray}
V_{jl}^{i} &=&\frac{d_{\alpha }^{2}}{(2\pi )^{2}\hbar c^{3}}%
\sum\limits_{l,j=1}^{N_{a}}\int\limits_{0}^{\infty }\omega _{k}^{3}d\omega
_{k}\int d\Omega _{k}\int\limits_{0}^{t}d\tau \exp [i(\omega _{i}-\omega
_{k})\tau ]  \nonumber \\
&&\times (1-(\mathbf{e}_{k},n_{d}))\langle \lbrack \tilde{J}%
_{j}^{+}(t),O(t)]\ \tilde{J}_{l}^{-}(t-\tau )\rangle \exp [i\omega
_{k}r_{jl}\cos \theta ],  \nonumber \\
i &\equiv &a,\ \ \ ,b.  \label{eq:int1}
\end{eqnarray}%
Here the frequency $\omega _{i}$ corresponds to $\mathit{A}$ and $\mathit{B}$%
\textit{\ }atomic systems; $i=r,s$; for $i=r$ operators $\tilde{J}_{j}^{+}$,$%
\ \ \tilde{J}_{l}^{-}$ corresponds to $\tilde{R}_{j}^{+}$ ,\ $\tilde{R}%
_{l}^{-}$ and for $i=s$ these operators corresponds to $\tilde{S}_{j}^{+}$,$%
\ $\ $\tilde{S}_{l}^{-}$\textbf{\ } Passing to new variable $\upsilon
=\omega _{k}-\omega _{i}$ and considering that the smooth function $\omega
_{k}$ under integral \ can be approximation with $\omega _{i}^{3}$, we
obtain the following approximate expression of thirst order exchange
integrals
\begin{eqnarray}
\omega _{i}^{3}\int\limits_{-\omega _{k}}^{\infty }dv\exp [iv(\tau
-r_{jl}\cos \theta /c)] &\eqsim &\omega _{i}^{3}\int\limits_{-\infty
}^{\infty }dv\exp [iv(\tau -r_{jl}\cos \theta /c)]  \nonumber \\
&=&2\pi \omega _{i}^{3}\delta (\tau -r_{jl}\cos \theta /c).
\label{eq:intaprox}
\end{eqnarray}%
$\omega _{i}$ is the emission frequency relatively the dipole active
transitions of the $R$ and $S$ atomic subsystems. In this approximation I
obtain the following integral on angle $\theta $ and retardation $\tau $%
\begin{eqnarray}
V_{jl}^{i} &=&\frac{\omega _{i}^{3}d_{i}^{2}}{2\hbar c^{3}}%
\int\limits_{0}^{\pi }d\theta \int\limits_{0}^{t}d\tau \sin \theta \delta
(\tau -r_{jl}\cos \theta /c)D_{i}(\theta )\langle \lbrack \tilde{J}%
_{j}^{+}(t),O(t)]\ \tilde{J}_{l}^{-}(t)\rangle  \nonumber \\
&=&\frac{\omega _{i}^{3}d_{i}^{2}}{2\hbar c^{3}}\int\limits_{0}^{\pi
}d\theta \sin \theta \Theta (\cos \theta )D_{jl}%
\bigl[%
\frac{\partial }{\partial \omega _{i}}%
\bigr]%
\exp [i\omega _{i}r_{jl}\cos \theta ]\langle \lbrack \tilde{J}%
_{j}^{+}(t),O(t)]\ \tilde{J}_{l}^{-}(t)\rangle  \nonumber \\
&=&\frac{1}{2\tau _{i}}\chi (j,l)\langle \lbrack \tilde{J}_{j}^{+}(t),O(t)]\
\tilde{J}_{l}^{-}(t)\rangle .  \label{eq:ex1}
\end{eqnarray}%
Here $\tau _{i}$ and $\chi (j,l)$ are the spontaneous emission time and
exchange integral between the single photon radiators respectively \cite%
{GH_1982}
\begin{equation}
\tau _{i}=\frac{3\hbar c^{3}}{4d_{i}\omega _{i}^{3}},\ \ \ \chi (j,l)=D_{jl}%
\bigl[%
\frac{\partial }{\partial \omega _{i}}%
\bigr]%
\frac{3}{4}\frac{\exp [i\omega _{i}r_{jl}/c]-1}{i\omega _{i}r_{jl}/c},
\label{eq:exp1}
\end{equation}%
the expressions in equation (\ref{eq:ex1}) and (\ref{eq:exp1}) are defined
below
\begin{eqnarray}
D_{i}(\theta ) &=&1+\cos ^{2}\xi _{i}-\cos ^{2}\theta (3\cos ^{2}\xi _{i}-1),
\nonumber \\
D_{jl}%
\bigl[%
\frac{\partial }{\partial \omega _{i}}%
\bigr]
&=&%
\bigl[%
1+\cos ^{2}(\xi _{i})+(3\cos ^{2}\xi _{i}-1)\frac{c^{2}}{r_{jl}^{2}}\frac{%
\partial ^{2}}{\partial \omega _{i}^{2}}]^{2}%
\bigr]
\label{eq:Dteta}
\end{eqnarray}%
where $\cos \xi _{i}$ is the scalar product between the unitary vectors
along the direction of dipole momentum of the $j$ (or $l$) atom $\mathbf{n}%
_{d_{i}}=\mathbf{d}_{i}/d_{i}$ and the direction of the distance between the
$j$ and $l$ atoms $\mathbf{n}_{jl}=\mathbf{r}_{jl}/r_{jl}$.

\textbf{2.} The two-photon exchange integral between dipole forbidden
transition of the radiators of $D$ subsystem is described by third term in
the right hand site of equation (\ref{eq:O2})
\begin{eqnarray}
V_{jl}^{b} &=&\frac{V^{2}}{(2\pi )^{6}}\int\limits_{0}^{2\pi }d\varphi
_{1}\int\limits_{0}^{\pi }d\theta _{1}\sin \theta
_{1}\int\limits_{0}^{\infty }k_{1}^{2}dk_{1}\int\limits_{0}^{2\pi }d\varphi
_{2}\int\limits_{0}^{\pi }d\theta _{2}\sin \theta
_{2}\int\limits_{0}^{\infty }k_{2}^{2}dk_{2}\frac{(\mathbf{n}_{eg},\mathbf{e}%
_{\lambda _{1}})^{2}(\mathbf{n}_{ei},\mathbf{e}_{\lambda
_{2}})^{2}q^{2}(\omega _{1},\omega _{2})}{\hbar ^{2}}  \nonumber \\
&\times &\int\limits_{0}^{t}d\tau \exp [-i(2\omega _{0}-\omega _{1}-\omega
_{2})\tau +i(\mathbf{k}_{1}+\mathbf{k}_{2},\mathbf{r}_{j}-\mathbf{r}%
_{l})]\left\langle [\tilde{D}_{j}(t),O(t)]\tilde{D}_{l}(t-\tau
)\right\rangle .  \label{eq:int2}
\end{eqnarray}%
The exchange integral in Born-Marcov approximation for two-photon emission
was obtained in paper \cite{E_1988}. Here we will estimate the exchange
integral of expression (\ref{eq:int2}) integration firstly on the wave
vectors $\mathbf{k}_{1}$ and $\mathbf{k}_{2}.$ Indeed considering that the
amplitude $q^{2}(\omega _{1},\omega _{2})$ is the smooth function of the
variables $k_{1}$ and $k_{2}$ in comparison with rapid oscillation functions
$\exp [i(\mathbf{k}_{1}+\mathbf{k}_{2},\mathbf{r}_{j}-\mathbf{r}_{l})]$ and $%
\exp [-i(2\omega _{0}-\omega _{k_{1}}-\omega _{k_{2}})\tau ]$, we can
approximate the amplitude $q^{2}(\omega _{1},\omega _{2})$ according with
the resonance frequencies of the atomic system $q^{2}(\omega _{1},\omega
_{2})\eqsim q^{2}(\omega _{0},\omega _{0})$. It is not difficult to observe
that the maximal value of the smooth function under the integral is obtained
for frequencies $\omega _{k_{2}}=\omega _{k_{1}}\eqsim \omega _{0}$. In
order to integrate this expression on the variables $k_{1}$ and $k_{2}$ we
change the variables $x_{i}=\omega _{k_{i}}-\omega _{i}$ as in expression (%
\ref{eq:intaprox}), and consider that $\tilde{D}_{j}^{+}(t)=D_{j}^{+}(t)\exp
[-2i\omega _{0}t]$ is smooth operator. In these approximations it is obtain
the following expression for $V_{jl}^{b}$
\begin{eqnarray}
V_{jl}^{b} &=&\frac{V^{2}q^{2}(\omega _{0},\omega _{0})}{(2)^{4}\pi ^{2}}%
\int\limits_{0}^{t}d\tau \int\limits_{0}^{\pi }d\theta _{1}\sin \theta
_{1}\int\limits_{0}^{\pi }d\theta _{2}\sin \theta _{2}D_{0}(\theta
_{1})D_{0}(\theta _{2})\exp [i\omega _{0}r_{jl}(\cos \theta _{1}+\cos \theta
_{2})/c]  \nonumber \\
&&\delta (\tau -r_{jl}\cos \theta _{1}/c)\delta (\tau -r_{jl}\cos \theta
_{2}/c)\left\langle [\tilde{D}_{j}(t),O(t)]\tilde{D}_{l}(t-\tau
)\right\rangle  \nonumber \\
&\eqsim &\frac{V^{2}q^{2}(\omega _{0},\omega _{0})}{(2)^{4}\pi ^{2}}%
D_{jl}^{2}\left( \frac{\partial }{\partial \omega _{0}}\right) \frac{\exp
[2i\omega _{0}r_{jl}]-1}{2i\omega _{0}r_{jl}}\left\langle [\tilde{D}%
_{j}(t),O(t)]\tilde{D}_{l}(t-\tau )\right\rangle ,  \label{eq:VB}
\end{eqnarray}%
where expressions $D_{0}(\theta _{1})$ and \ $D_{jl}(\partial /(\partial
\omega _{0})$ are defined by the expressions (\ref{eq:Dteta}). Integrating
the right hand site of the equation (\ref{eq:VB}) on the solid angle and
retardation, we obtain the following approximative expression%
\[
V_{jl}^{b}=\frac{1}{2\tau _{b}}\chi _{b}(j,l)\left\langle [\tilde{D}%
_{j}^{+}(t),O(t)]\tilde{D}_{l}^{-}(t)\right\rangle ,
\]%
where%
\begin{eqnarray}
\frac{1}{2\tau _{b}} &=&\frac{2^{2}}{3^{2}}\frac{\omega
_{0}^{7}d_{23}^{2}d_{31}^{2}}{2^{2}\pi \hbar ^{2}c^{6}}\{3/2\}\left\{ \frac{1%
}{\omega _{32}+\omega _{0}}+\frac{1}{\omega _{31}-\omega _{0}}\right\} ^{2},
\nonumber \\
\chi _{b}(j,l) &=&\frac{3^{2}}{4}\frac{\pi c}{4\omega _{0}r_{jl}}D^{2}%
\bigl[%
\frac{\partial }{\partial \omega _{0}}%
\bigr]%
\frac{\exp [2i\omega _{0}r_{jl}/c]-1}{i\omega _{0}r_{jl}/c}  \label{eq:exp2}
\end{eqnarray}%
This exchange integral diverges, when the distance between the radiators $%
r_{jl}$ is less than the wavelength $\lambda _{0}=2\pi c/\omega _{0}$. In
order to take in to account the value of the exchange integral for the small
parameter, $r_{jl}/\lambda _{0}<<1,$ let us integrate the expression (\ref%
{eq:int2}) taking in to account the method proposed in papers \cite{E_1988}
and \cite{EM_1997}. In this case we obtain the following expression for $%
ReV_{jl}^{b}$%
\begin{eqnarray*}
F(j,l) &=&ReV_{jl}^{b}=\frac{d_{31}^{2}d_{32}^{2}}{4\pi \hbar ^{2}c^{6}}%
\int\limits_{0}^{2\omega _{0}}d\omega _{k}\omega _{k}^{3}(\omega
_{21}-\omega _{k})^{3} \\
&\times &\chi _{jl}(\omega _{k})\chi _{jl}(\omega _{21}-\omega _{k})\left\{
\frac{1}{\omega _{31}-\omega _{k_{1}}}+\frac{1}{\omega _{32}+\omega _{k_{1}}}%
\right\} ^{2},
\end{eqnarray*}%
where
\[
\chi _{jl}(\omega )=(1-\cos ^{2}{\xi })\frac{\sin {\frac{\omega r_{jl}}{c}}}{%
\frac{\omega r_{jl}}{c}}+(1-3\cos ^{2}{\xi })\left\{ \frac{\cos {\frac{%
\omega r_{jl}}{c}}}{(\frac{\omega r_{jl}}{c})^{2}}-\frac{\sin {\frac{\omega
r_{jl}}{c}}}{(\frac{\omega r_{jl}}{c})^{3}}\right\} .
\]

\textbf{3. }In the right hand part of this equation (\ref{eq:O2}) the third
order terms contains the resonances between dipole active radiators $\mathit{%
A}$\textit{\ ,}$\mathit{B}$ and dipole forbidden radiators $\mathit{D}$%
\textit{,} described by the correlation functions $\langle
S_{l}^{+}(t)[R_{j}^{+}(t),O(t)]D_{m}^{-}(t)\rangle $, $\langle \lbrack
D_{n}^{+}(t),O(t)]R_{j}^{-}(t)S_{l}^{-}(t)\rangle $ and $\langle
R_{l}^{+}(t)[S_{j}^{+}(t),O(t)]D_{m}^{-}(t)\rangle $. Let us introduced the
similar approximation in the chronological interaction between the atomic
subsystems $\mathit{A}$, $\mathit{B}$, and $\mathit{D}$. The exchange
integral between three atoms is represented in the similar form as in the
expression (\ref{eq:int2})

\begin{eqnarray}
V_{jln;as-d}^{c} &=&i\frac{V}{(2\pi )^{3}}\frac{V}{(2\pi )^{3}}\frac{2\pi
\hbar d_{s}d_{r}}{Vc^{6}\hbar ^{3}}\frac{(2\pi )^{3}}{V}(2\pi )^{2}\left(
\frac{1}{2}\right) ^{2}\int\limits_{0}^{\infty }\omega _{1}^{2}d\omega
_{1}\int\limits_{0}^{\infty }\omega _{2}^{2}d\omega _{2}\sqrt{\omega
_{1}\omega _{2}}\chi (\omega _{1},\omega _{2})  \nonumber \\
&\times
&\int\limits_{-1}^{1}dx_{1}\int\limits_{-1}^{1}dx_{2}\int\limits_{0}^{t}d%
\tau _{1}\int\limits_{0}^{t}d\tau _{2}\exp [-i(\omega _{1}-\omega _{r})\tau
_{1}-i(\omega _{2}-\omega _{s})\tau _{2}]  \nonumber \\
&\times &D_{nl}%
\bigl[%
\frac{\partial }{\partial \omega _{s}}%
\bigr]%
D_{nj}%
\bigl[%
\frac{\partial }{\partial \omega _{r}}%
\bigr]%
\exp [i\omega _{1}r_{nj}x_{1}/c+i\omega _{2}r_{nl}x_{2}/c]  \nonumber \\
&&\langle \lbrack \tilde{D}_{n}^{+}(t),O(t)]\tilde{R}_{j}^{-}(t-\tau _{1})%
\tilde{S}_{l}^{-}(t-\tau _{2})\rangle ;  \label{eq:int3}
\end{eqnarray}%
After the substitution of variables $\omega _{1}-\omega _{r}=\tilde{\omega}%
_{1}$ and $\omega _{2}-\omega _{s}=\tilde{\omega}_{2}$, we can approximate
the smooth amplitude $\omega _{1}^{2}\omega _{2}^{2}\sqrt{\omega _{1}\omega
_{2}}\chi (\omega _{1},\omega _{2})$ with expression $(\omega
_{r})^{2}\omega _{s}^{2}\sqrt{\omega _{r}\omega _{s}}\chi (\omega
_{r},\omega _{s})$. Integrals on the new variable $\tilde{\omega}_{1}$ and $%
\tilde{\omega}_{1}$ give the following aspect of expression (\ref{eq:int3})%
\begin{eqnarray*}
V_{jln;as-d}^{c} &=&2\pi i\left( \frac{1}{2}\right) ^{2}\frac{d_{s}d_{r}}{%
c^{6}\hbar ^{2}}\omega _{s}^{2}(\omega _{r})^{2}\sqrt{\omega _{r}\omega _{s}}%
\chi (\omega _{s},\omega
_{r})\int\limits_{0}^{1}dx_{1}\int\limits_{0}^{1}dx_{2} \\
&\times &D_{nl}%
\bigl[%
\frac{\partial }{\partial \omega _{s}}%
\bigr]%
D_{nj}%
\bigl[%
\frac{\partial }{\partial \omega _{r}}%
\bigr]%
\exp [i\omega _{r}r_{nj}x_{1}/c+i\omega _{s}r_{nl}x_{2}/c] \\
&\times &\langle \lbrack \tilde{D}_{n}^{+}(t),O(t)]\tilde{R}%
_{j}^{-}(t-r_{nj}x_{1}/c)\tilde{S}_{l}^{-}(t-r_{nl}x_{2}/c)\rangle
\end{eqnarray*}%
from which follows that $x_{1},$and $x_{2}>0$ . In the Born approximation
the expression for $V_{jln;as-d}^{c}$
\[
V_{rs-d}(m,j,l)=\frac{i}{4\tau _{bsr}}U(j,l,m)\langle \lbrack \tilde{D}%
_{n}^{+}(t),O(t)]\tilde{R}_{j}^{-}(t)\tilde{S}_{l}^{-}(t)\rangle
\]%
where
\begin{eqnarray}
\frac{1}{\tau _{bsr}} &=&\left( \frac{2}{3}\right) ^{2}\frac{%
d_{s}d_{r}d_{23}d_{31}\omega _{s}^{3}(\omega _{r})^{3}}{4\pi c^{6}\hbar ^{2}}%
\left\{ \frac{1}{\omega _{32}+\omega _{s}}+\frac{1}{\omega _{31}-\omega _{r}}%
\right\} ,  \nonumber \\
U(j,l,m) &=&-\left( \frac{3}{2}\right) ^{2}D_{nl}%
\bigl[%
\frac{\partial }{\partial \omega _{s}}%
\bigr]%
D_{nj}%
\bigl[%
\frac{\partial }{\partial \omega _{r}}%
\bigr]%
\frac{c^{2}[\exp i\omega _{r}r_{nj}/c]-1][\exp [i\omega _{s}r_{nl}/c]-1]}{%
\omega _{r}\omega _{s}r_{nj}r_{nl}}.  \label{eq:U}
\end{eqnarray}%
\textbf{4.} Let now found the retardation in the last correlation function
term of equation (\ref{eq:O2}).Taking in to account the retardation in the
rapid oscillation part of atomic operators $J^{\pm }(t-\tau )=\tilde{J}^{\pm
}(t-\tau )\exp [\pm i\omega (t-\tau )]$ where $J^{\pm }$, $\omega $ and $%
\tau $ are the atomic operators, transition frequencies and delay time for
atomic subsystems $\mathit{A}$\textit{, }$\mathit{B}$ and $\mathit{D}$%
\textit{\ }respectively,

\begin{eqnarray}
V_{;ab-d}(t)
&=&i\sum\limits_{k_{1}k_{2}}\sum\limits_{m=1}^{N}\sum\limits_{l=1}^{N_{a}}%
\sum\limits_{j=0}^{N_{b}}\frac{(\mathbf{d}_{r},\mathbf{g}_{k_{1}})(\mathbf{d}%
_{s},\mathbf{g}_{k_{2}})(\mathbf{n}_{eg},\mathbf{e}_{\lambda _{1}})(\mathbf{n%
}_{ei},\mathbf{e}_{\lambda _{2}})q(\omega _{k_{1}},\omega _{k_{2}})}{\hbar
^{3}}  \nonumber \\
&\times &\int\limits_{0}^{t}d\tau _{1}\exp [i(2\omega _{0}-\omega
_{k_{1}}-\omega _{k_{2}})\tau _{1}]\int\limits_{0}^{t}d\tau _{2}\exp
[-i(\omega _{s}-\omega _{k_{1}})\tau _{2}]  \nonumber \\
&\times &\exp [-i(\mathbf{k}_{1},\mathbf{r}_{j}-\mathbf{r}_{m})+i(\mathbf{k}%
_{2},\mathbf{r}_{l}-\mathbf{r}_{m})]\langle \tilde{S}_{l}^{+}(t-\tau _{2})[%
\tilde{R}_{j}^{+}(t),O(t)]\tilde{D}_{m}^{-}(t-\tau _{1})\rangle .
\label{ex:Vjlm}
\end{eqnarray}%
Passing from the summation to integration in expression \ (\ref{ex:Vjlm})\
we obtain following expression for correlation between the $j$, $l$ and $m$
atoms

\begin{eqnarray}
V_{rs-d}(m,j,l) &=&i\frac{V}{(2\pi )^{3}}\frac{V}{(2\pi )^{3}}\frac{2\pi
\hbar d_{s}d_{r}}{Vc^{6}\hbar ^{3}}\frac{(2\pi )^{3}}{V}(2\pi )^{2}\left(
\frac{1}{2}\right) ^{2}\int\limits_{0}^{\infty }\omega _{1}^{2}d\omega
_{1}\int\limits_{0}^{\infty }\omega _{2}^{2}d\omega _{2}  \nonumber \\
&&\times \sqrt{\omega _{1}\omega _{2}}\chi (\omega _{1},\omega
_{2})\int\limits_{-1}^{1}dx_{1}\int\limits_{-1}^{1}dx_{2}\int%
\limits_{0}^{t}d\tau _{1}\exp [i(2\omega _{0}-\omega _{1}-\omega _{2})\tau
_{1}]  \nonumber \\
&&\times \int\limits_{0}^{t}d\tau _{2}\exp [-i(\omega _{s}-\omega _{2})\tau
_{2}]  \nonumber \\
&&\times D_{jl}%
\bigl[%
\frac{\partial }{\partial \omega _{2}}%
\bigr]%
D_{jm}%
\bigl[%
\frac{\partial }{\partial \omega _{1}}%
\bigr]%
\exp [-i\omega _{2}r_{ml}x_{2}/c-i\omega _{1}r_{jm}x_{1}/c)]  \nonumber \\
&&\times \langle \tilde{S}_{l}^{+}(t-\tau _{2})[\tilde{R}_{j}^{+}(t),O(t)]%
\tilde{D}_{m}^{-}(t-\tau _{1}).  \label{eq:int4}
\end{eqnarray}%
Introducing the new variables $u_{1}=\omega _{1}-\omega _{r}$; $u_{2}=\omega
_{2}-\omega _{s}$ in (\ref{eq:int4}), and approximating the smooth amplitude
$\omega _{1}^{2}\omega _{2}^{2}\sqrt{\omega _{1}\omega _{2}}\chi (\omega
_{1},\omega _{2})$ with expression $\omega _{s}^{2}\omega _{r}^{2}\sqrt{%
\omega _{s}\omega _{r}}\chi (\omega _{s},\omega _{r})$ we get using the
delta functions (\ref{eq:intaprox})%
\begin{eqnarray*}
V_{jlm;as-d}(j,l) &=&2\pi i\left( \frac{1}{2}\right) ^{2}\frac{d_{s}d_{r}}{%
c^{6}\hbar ^{2}}\omega _{s}^{2}\omega _{r}^{2}\sqrt{\omega _{s}\omega _{r}}%
\chi (\omega _{s},\omega
_{r})\int\limits_{-1}^{1}dx_{1}\int\limits_{-1}^{1}dx_{2}\int%
\limits_{0}^{t}d\tau _{1}\int\limits_{0}^{t}d\tau _{2} \\
&\times &\delta (\tau _{2}-\tau _{1}-r_{jl}x_{2}/c)\delta (\tau
_{1}-r_{jm}x_{1}/c) \\
&&\times D_{jl}%
\bigl[%
\frac{\partial }{\partial \omega _{s}}%
\bigr]%
D_{jm}%
\bigl[%
\frac{\partial }{\partial \omega _{r}}%
\bigr]%
\exp [i\omega _{s}r_{ml}x_{2}/c]\exp [-i\omega _{r}r_{jm}x_{1}/c] \\
&&\times \langle \tilde{S}_{l}^{+}(t-\tau _{2})[\tilde{R}_{j}^{+}(t),O(t)]%
\tilde{D}_{m}^{-}(t-\tau _{1})\rangle ,
\end{eqnarray*}%
the value of which can be estimated observing from the arguments of $\delta $%
-functions that $x_{1}>0$ and $\ r_{jm}x_{1}+r_{jl}x_{2}>0$.\ In this case,
neglecting the retardation $\tau _{1}$ and $\tau _{2}$ in the smooth
correlation function \ we obtain
\[
V_{jlm;as-d}(j,l)=\frac{i}{4\tau _{sbr}}V(j,l.m)\langle \tilde{S}_{l}^{+}(t)[%
\tilde{R}_{j}^{+}(t),O(t)]\tilde{D}_{m}^{-}(t)\rangle ,
\]%
where cooperative rate is
\[
\frac{1}{\tau _{sbr}}=\left( \frac{2}{3}\right) ^{2}\frac{%
d_{s}d_{r}d_{23}d_{31}\omega _{s}^{3}\omega _{r}^{3}}{4\pi c^{6}\hbar ^{2}}%
\left\{ \frac{1}{\omega _{32}+\omega _{s}}+\frac{1}{\omega _{31}-\omega _{r}}%
\right\} .
\]%
the integral $V(j,l.m)$ on the direction of the emitted photons is
\begin{eqnarray}
V(j,l.m) &=&\left( \frac{3}{2}\right) ^{2}D_{ml}%
\bigl[%
\frac{\partial }{\partial \omega _{s}}%
\bigr]%
D_{jm}%
\bigl[%
\frac{\partial }{\partial \omega _{r}}%
\bigr]
\nonumber \\
\times &&\bigl[\frac{c^{2}\{2\exp [i(\omega _{s}r_{ml}-\omega
_{r}r_{jm})/c]-2\exp [2i\omega _{s}r_{ml}]-\exp [-2i\omega _{r}r_{jm}]+1]\}}{%
2\omega _{s}\omega _{r}r_{lm}r_{ml}}\theta (r_{ml}-r_{mj})  \nonumber \\
+. &&%
\bigl(%
\frac{c^{2}(\omega _{s}+\omega _{r})[\exp (-i\omega _{r}r_{mj}/c)-1]\exp
(i\omega _{s}r_{ml}/c)-\exp [-i\omega _{s}r_{ml}/c-i\omega _{r}r_{mj}/c]}{%
(\omega _{s}+\omega _{r})\omega _{s}\omega _{r}r_{mj}r_{ml}}  \nonumber \\
+ &&\frac{c^{2}[\omega _{s}\exp [-i(\omega _{s}+\omega _{r})r_{ml}/c]+\omega
_{r}]}{(\omega _{s}+\omega _{r})\omega _{r}\omega _{s}r_{jm}r_{ml}}%
\bigr)%
\bigr]%
\theta (r_{mj}-r_{jl}).  \label{eq:Vjlm}
\end{eqnarray}%
If we will consider that $x_{1}>0,$ and $x_{2}>0$ the expression for (\ref%
{eq:Vjlm}) takes more simple form
\begin{eqnarray*}
V(j,l.m) &\eqsim &\left( \frac{3}{2}\right) ^{2}D_{ml}%
\bigl[%
\frac{\partial }{\partial \omega _{s}}%
\bigr]%
D_{jm}%
\bigl[%
\frac{\partial }{\partial \omega _{r}}%
\bigr]
\\
&&\frac{c^{2}\{\exp [i\omega _{s}r_{ml}/c]-1][\exp [-i\omega _{r}r_{jm}/c]-1]%
}{\omega _{s}\omega _{r}r_{lm}r_{ml}}
\end{eqnarray*}
The expression for exchange integrals described in points $1.-4.$ are used
in the master equation (\ref{eq:O4})

\end{document}